\documentclass[11pt]{article}

\usepackage{amstext,amsmath,amssymb,amsfonts,bbm}
\usepackage{hyperref}
\usepackage{authblk}
\usepackage{caption}
\usepackage[latin1]{inputenc}
\usepackage{epsfig}
\usepackage{subfigure}
\usepackage{color}
\usepackage{graphicx}
\usepackage{slashed}

\usepackage[lmargin=80pt,rmargin=80pt,tmargin=80pt,bmargin=80pt]{geometry}

\newcommand{\be}{\begin{equation}}
\newcommand{\ee}{\end{equation}}
\newcommand{\nn}{\nonumber}
\newcommand{\f}{\frac}
\newcommand{\p}{\partial}

\newcommand{\Tr}{{\rm Tr}}

\newcommand{\la}{\langle}
\newcommand{\ra}{\rangle}

\def\extd{\mathrm {d}}

\newcommand{\one}{\mbox{$1 \hspace{-1.0mm}  {\bf l}$}}

\DeclareMathOperator{\im}{\mathrm{i}}

\let\a=\alpha \let\b=\beta  \let\g=\gamma  \let\d=\delta
       \let\k=\kappa \let\l=\lambda
\let\m=\mu    \let\n=\nu          \let\r=\rho \let\om=\omega
\let\s=\sigma      
    
\let\G=\Gamma    \let\L=\Lambda 
         
  \let\eps=\epsilon

\newcommand{\psib}{\bar{\psi}}

\newcommand{\cA}{\mathcal{A}}

\newcommand{\cD}{\mathcal{D}}

\newcommand{\cI}{\mathcal{I}}
\newcommand{\cJ}{\mathcal{J}}
\newcommand{\cL}{\mathcal{L}}
\newcommand{\cM}{\mathcal{M}}

\newcommand{\cO}{\mathcal{O}}

\newcommand{\cS}{\mathcal{S}}

\begin{document}

\title{\bf Tensorial Gross-Neveu models}

\author[1]{Dario Benedetti}
\author[2]{Sylvain Carrozza}
\author[3]{Razvan Gurau}
\author[4]{Alessandro Sfondrini}

\affil[1]{\normalsize\it Laboratoire de Physique Th\'eorique (UMR8627), CNRS, Univ.Paris-Sud, \authorcr
\it Universit\'e Paris-Saclay, 91405 Orsay, France \authorcr
email: dario.benedetti@th.u-psud.fr  \authorcr \hfill }

\affil[2]{\normalsize\it Perimeter Institute for Theoretical Physics \authorcr \it
 31 Caroline St N, Waterloo, ON N2L 2Y5, Canada  \authorcr email: scarrozza@perimeterinstitute.ca  \authorcr \hfill}

\affil[3]{\normalsize\it Centre de Physique Th\'{e}orique, \'{E}cole
Polytechnique, CNRS, F-91128 Palaiseau, France and
Perimeter Institute for Theoretical Physics, 31 Caroline St. N, N2L 2Y5, Waterloo, ON,
Canada \authorcr email: rgurau@cpht.polytechnique.fr \authorcr \hfill}

\affil[4]{\normalsize\it Institut f\"ur theoretische Physik, ETH Z\"urich,
Wolfgang-Pauli-Stra{\ss}e 27, 8093 Z\"urich, Switzerland \authorcr email: sfondria@itp.phys.ethz.ch}

\date{}

\maketitle

\hrule\bigskip

\begin{abstract}

We define and study various tensorial generalizations of the Gross-Neveu model in two dimensions, that is, models with four-fermion interactions and $G^3$ symmetry, where we take either $G=U(N)$ or $G=O(N)$.
Such models can also be viewed as two-dimensional generalizations of the Sachdev-Ye-Kitaev model, or more precisely of its tensorial counterpart introduced by  Klebanov and Tarnopolsky, which is in part our motivation for studying them.
Using the Schwinger-Dyson equations at large-$N$, we discuss the phenomenon of dynamical mass generation and possible combinations of couplings to avoid it.
For the case  $G=U(N)$, we introduce an intermediate field representation and perform a stability analysis of the vacua. 
It turns out that the only apparently viable combination of couplings that avoids mass generation corresponds to an unstable vacuum. 
The stable vacuum breaks $U(N)^3$ invariance, in contradiction with the Coleman-Mermin-Wagner theorem, but this is an artifact of the large-$N$ expansion, similar to the breaking of continuous chiral symmetry in the chiral Gross-Neveu model.

\end{abstract}

\hrule\bigskip

\newpage
\tableofcontents

\section{Introduction}

The large-$N$ limit has a long tradition in field theory as it allows to restrict the perturbative expansion to a specific class of Feynman diagrams. In the case of vector fields with $N$ components, the restricted class of diagrams is so simple 
(essentially just chains of bubbles diagrams) that it generally allows to solve many models, either diagrammatically or by saddle point methods \cite{Moshe:2003xn}. For the next natural generalization in the sense of linear algebra, 
the case of $N\times N$ matrix fields, the large-$N$ limit has led to numerous results in zero dimensions, in particular because of its connection to two-dimensional quantum gravity and string theory \cite{DiFrancesco:1993cyw}. However, 
matrix field theories in two or higher dimensions turn out to be much more complicated than the vector case, as at leading order in the large $N$ expansion one encounters all the planar diagrams. Somewhat surprisingly, it turns out that going one more step up in the rank, i.e. considering tensor fields, things simplify again, although not to the level of vector fields, thus making tensor fields the good candidates 
for models with a new and manageable, yet non-trivial, large-$N$ limit. 

Although tensor models in zero dimensions were introduced long ago as a generalization of matrix models for higher-dimensional quantum gravity \cite{Ambjorn:1990ge,Sasakura:1990fs}, 
it is only more recently that the possibility of studying their $1/N$ expansion was discovered \cite{Gurau:2010ba,Gurau:2011aq,Gurau:2011xq}, and the class of leading order diagrams 
identified: the melons \cite{Bonzom:2011zz,Bonzom:2012hw}. The latter are a class of diagrams with a tree-like combinatorial structure; they are therefore easier to sum than the planar diagrams, but 
are definitely richer than the chains of bubbles. Such combinatorial structures show up naturally in a variety of other contexts, and have for instance played a central role in renormalization group approaches to group field theories \cite{Oriti:2011jm, BGR, COR, CarrozzaThesis, BBO, Benedetti:2015yaa, Carrozza:2016vsq}. 
More recently, it is precisely the structure of the melonic graphs that has led to the observation that the Sachdev-Ye-Kitaev (SYK) 
model \cite{SachdevYe, Kitaev2015, MaldacenaStanford, PolchinskiRosenhaus} and its tensorial generalizations \cite{WittenTensor,KlebanovTarnopolsky,Krishnan:2017ztz, Bonzom:2017pqs,Bulycheva:2017ilt,Choudhury:2017tax,Peng:2016mxj} have a conformal IR limit 
and saturate the chaos bound \cite{MSS}, thus establishing them as interesting models of holography. Such models live in one (time) dimension, but have also been 
generalized to higher dimensions \cite{KlebanovTarnopolsky, KGT, Berkooz:2017efq, Murugan:2017eto}. 
In close relation to such developments, a new expansion has recently been proposed to investigate the properties of matrix models of quantum black holes in large $D$ dimensions 
\cite{Ferrari:2017ryl, Azeyanagi:2017drg, Ferrari:2017jgw, Azeyanagi:2017mre}.  

In this paper, we explore two-dimensional models of fermions in a tensor representation of rank three, with quartic interactions. These are a natural generalization of the Gross-Neveu (GN) model \cite{Gross:1974jv}, 
in which the fermions are in a vector representation. The GN model is asymptotically free in two dimensions, and although a mass term is forbidden by discrete chiral symmetry, dynamical symmetry breaking occurs and a mass 
is non-perturbatively generated. We expect a similar behavior in the tensorial generalization, thus in general we do not expect to find a conformal field theory in the IR, which is the main attractive feature of SYK-like models.
However, the tensorial generalizations have several couplings and therefore it is in principle possible that for some specific values of the couplings a mass is \emph{not} generated and a conformal theory is found. 
We will investigate this possibility by analyzing the large-$N$ Schwinger-Dyson equations first, and then by the intermediate field method.

The two-point function of SYK models enjoys an emergent time reparametrization invariance (or conformal invariance) in the large $N$ and strong coupling limit, apparent at the level of its Schwinger-Dyson equation. 
A crucial ingredient is that the self-energy dominates over the free propagator in the infrared regime. For massless fermions (with standard $1/\slashed{p}$ propagator), a naive dimensional analysis 
shows that this property can only hold if $d(1- 2/q) < 1$, where $d$ is the spacetime dimension and $q$ is the order of the interactions. Hence $d=2$ is the critical dimension at which this condition 
starts failing\footnote{For massless bosons the critical dimension is instead $d=3$. Incidentally, the presence of non-trivial infrared fixed points in purely bosonic generalizations of SYK models have 
almost completely been excluded, except in $d=3 - \varepsilon$ \cite{KGT}.}: the quartic couplings being dimensionless, both the free propagator and the self-energy must be taken into account in the 
two-point Schwinger-Dyson equation. This is the most crucial difference between Gross-Neveu tensor models and SYK-like theories, 
leading to substantial qualitative differences in their large $N$ behaviors.

The plan of the paper is as follows. In section \ref{sec:models} we introduce our two models -- a $U(N)^3$ model with Dirac 
fermions and an $O(N)^3$ model with Majorana fermions -- paying special attention to the parametrization of their action in terms of
invariants. In section \ref{sec:sde}, we study the large-$N$ (melonic) Schwinger-Dyson equations for the two-point functions of both models, and provide a first analysis 
of the mechanism underlying the spontaneous generation of mass. In the Dirac model, we find a specific combination of the coupling constants which seems to ensure that the theory remains massless. 
In the Majorana model, we find also a candidate massless theory, but with a marginally irrelevant coupling constant.
Relying on the introduction of a matrix intermediate field for the Dirac model, we move on to an in-depth analysis of the would-be massless vacuum in 
section \ref{sec:Dirac-1color}. The effective potential of the intermediate field is computed, and shown to develop a symmetry breaking pattern in the regime of interest. 
This results in the instability of the would-be massless vacuum, triggered by a non-zero expectation value of the intermediate field. We discuss in detail the apparent symmetry 
breaking of the $U(N)^3$ symmetry, and derive the effective non-linear dynamics governing the associated massless modes. We recall how the latter should not be understood as 
Goldstone bosons since any apparently broken continuous symmetry must be restored in $d=2$ \cite{Mermin:1966fe, Coleman:1973ci}, and we describe the mechanism of symmetry 
restoration at play in this model. We finally close with a summary and a few concluding remarks in section \ref{sec:conclu}. 

\noindent\emph{\bf Note added:} 

\noindent During completion of this work a paper by Prakash and Sinha \cite{Prakash:2017hwq} has appeared, in which they also study tensorial fermionic 
models in more than one dimensions.  They consider a model for complex fermions but with symmetry group $U(N)\times O(N)\times U(N)$ which allows one to write the equivalent 
of our $I_2$ (or tetrahedral) interaction. They do not consider the equivalent of our $I_0$ and $I_1$ interactions, and they focus on $d\neq 2$, searching for a behavior similar to the one in 
the SYK model, where the free propagator term in the Schwinger-Dyson equations can be discarded in the IR limit. 
Although the two works are thus in most part complementary, we make a connection between the two in section \ref{sec:SDE-M}, where, relying on the RG analysis of appendix \ref{app:beta2}, we provide a check of their conjecture that the models with tetrahedral interaction have a weakly interacting IR fixed point in $2-\epsilon$ dimensions.

\section{The models}
\label{sec:models}

The Gross-Neveu (GN) model~\cite{Gross:1974jv} is a model of $N$  massless Dirac (or Majorana) fermions\footnote{See appendix \ref{app:conventions} for notations and conventions.} arranged into a vector~$\psi_i$, with $i=1\ldots N$, and invariant under $U(N)$ (or $O(N)$)\footnote{As noted in \cite{Dashen:1975xh}, in the Dirac case the model is also symmetric under $O(2N)$, of which $U(N)$ is a subgroup. This is seen by rewiting $\psi_i= \psi_1+\im \psi_2$, with real (Majorana) spinors~$\psi_1$ and~$\psi_2$. Therefore, the two versions of the GN model are essentially equivalent.} transformations $\psi_i\to U_{ij}\psi_j$, with $U\in U(N)$ (or $O(N)$) and summation on repeated indices
\be
S_{\rm GN} = \int d^d x \; \left( \psib_i  \slashed{\p} \psi_i -g (\psib_i \psi_i)^2 \right) \;.
\ee
The model is renormalizable and asymptotically free in $d=2$ for $g>0$, and it has a discrete chiral symmetry that (if unbroken) protects it from the generation of a mass term. A mass gap is however generated non-perturbatively, as most easily seen in the large-$N$ limit. A large-$N$ expansion is obtained by redefining $g=\l/N$ and keeping $\l$ finite.
Then the beta function at leading order in $1/N$ is
\be \label{eq:GN-fullBeta}
\b(\l) = - \f{2}{\pi} \l^2\, .
\ee
Since in the large-$N$ limit such a beta function is valid also at large $\l$, one encounters a Landau pole in the running coupling. The latter is interpreted as the appearance of a tachyon due to an instability of the invariant vacuum. An analysis of the effective potential for the intermediate (or Hubbard-Stratonovich) field reveals that the stable vacuum breaks discrete chiral symmetry and leads to a dynamical mass generation. 

Here we will study generalizations of the GN model, where we have $N^r$ fermions arranged into a rank-$r$ tensor $\psi_{i_1 \ldots i_r}$, such that the action is invariant under $U(N)^r$ (or $O(N)^r$ for Majorana fermions) transformations 
\be \label{eq:flavor-sym}
\psi_{i_1 \ldots i_r}\to U^{(1)}_{i_1 j_1} \ldots U^{(r)}_{i_r j_r}\,\psi_{j_1 \ldots j_r}\,,
\qquad
\text{with}
\quad
 U^{(1)},\ldots U^{(r)} \in U(N) \; (\text{or }O(N)) \;.
\ee
There are many possible invariant actions; here we restrict to the case of $d=2$ and quartic interactions, in order to have power-counting renormalizability. For simplicity we will restrict also to~$r=3$.

We will refer to the invariance \eqref{eq:flavor-sym} as {\it flavor symmetry}, and we will also introduce a discrete permutation symmetry, which we will refer to as  {\it color symmetry} \cite{Bonzom:2012hw, Gurau:book}.
Lastly, like it is done for the usual GN model, we will introduce also a discrete chiral symmetry
\be \label{eq:dct}
\psi \to \im \g_5 \psi\;, \;\;\; \psib \to  \im \psib \g_5 \;,
\ee
as well as its continuous counterpart, which we define in appendix~\ref{app:conventions}, see \eqref{eq:cct1} there. 

We split the action of our model  into a free and an interacting part:
\be
S =  S_{\rm free} + S_{\rm int} \;,
\ee
where $S_{\rm free}$ is bilinear in the spinor fields, while $S_{\rm int}$ is quadrilinear.
The construction of bilinear and quadrilinear terms which are invariant under Euclidean and chiral symmetries is recalled in appendix~\ref{app:invariants}. We will see below how to construct \textit{flavor-} and \textit{color-invariant} interactions.
Since the choice of Majorana versus Dirac fermions---and hence  $O(N)^3$ versus  $U(N)^3$ symmetry---leads to a slightly different set of interactions, as well as to a different normalization for the quadratic term, we shall discuss the two cases separately starting from Dirac fermions.

\subsection{Dirac fermions}

In the case of complex Dirac fermions we define the free part of the action as
\be \label{eq:free-D}
S_{\rm free} =   \int d^2 x \; \psib_{abc} \slashed{\p} \psi_{abc}  \;.
\ee
The discrete chiral symmetry \eqref{eq:dct} forbids a mass term in the action, and thus \eqref{eq:free-D} is the only bilinear invariant under all our symmetries. In GN a mass is dynamically generated at the non-perturbative level, and we will address in some detail the question of dynamical symmetry breaking for our models in the upcoming sections.

\begin{figure}[ht]
 \centering
        \begin{minipage}{0.25\textwidth}
            \centering 
             \includegraphics[scale = 0.75]{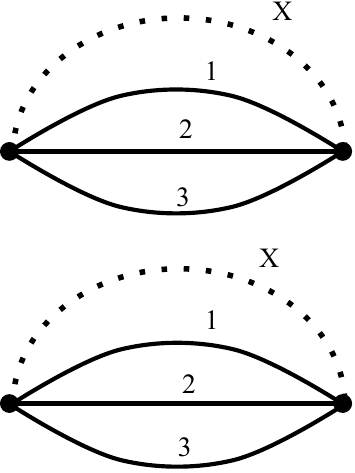}
        \end{minipage}
        \hspace{0.01\textwidth}
        \begin{minipage}{0.25\textwidth}
            \centering
            \includegraphics[scale = 0.75]{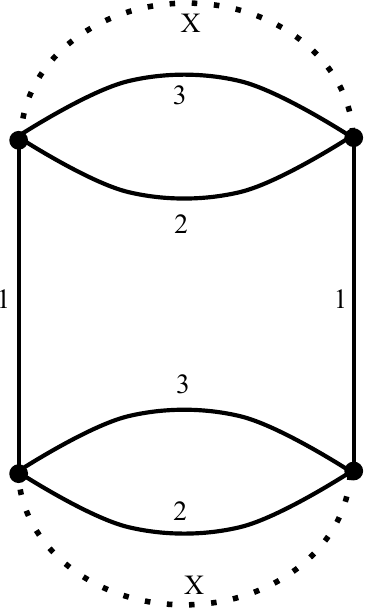}
        \end{minipage}
        \hspace{0.01\textwidth}
        \begin{minipage}{0.25\textwidth}
            \centering
            \includegraphics[scale = 0.75]{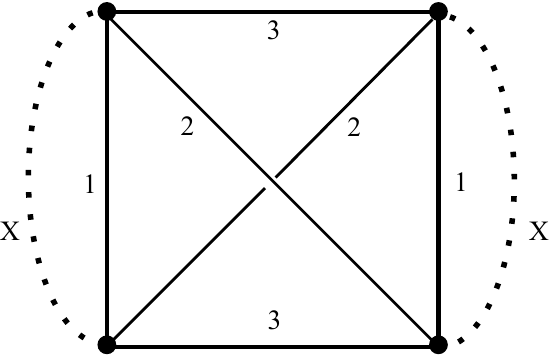}
        \end{minipage}
  \caption{\small{Graphical representation of the interaction vertices \eqref{eq:vertex0} (left), \eqref{eq:vertex1} (center), and \eqref{eq:vertex2} (right). Each vertex represents a tensor/spinor field, a solid line with label (color) $i\in\{1,2,3\}$ joining two vertices represents the contraction of the $i$-th  tensorial indices of the two fields at its end-vertices, and a dotted line with label $X$ represents the contraction of the spinorial indices of the fields at  its end-vertices, with the insertion of a matrix $\G^X$. 
  }}
\label{fig:vertices}
\end{figure}

$U(N)^3$-invariant interactions are constructed by pairwise contraction of indices belonging to $\psi$ fields with indices belonging to $\psib$ fields (which transform as the complex conjugate of \eqref{eq:flavor-sym}, where $U^{(i)}\in U(N)$).
In order to represent the interaction terms in a compact way it is convenient and customary to introduce a graphical representation, as explained in figure~\ref{fig:vertices}, where in the left and central panels the following interaction terms are represented:
\begin{align}
 \label{eq:vertex0}
 I_0^X = &  (\psib_{a_1 a_2 a_3} \Gamma^X \psi_{a_1 a_2 a_3})( \psib_{b_1 b_2 b_3} \Gamma^X \psi_{b_1 b_2 b_3} )\;,\\
 \label{eq:vertex1}
\begin{split}
I_1^{X,1} = 
& (\psib_{a_1a_2 a_3} \Gamma^X \psi_{b_1 a_2 a_3} )(\psib_{b_1 b_2 b_3} \Gamma^X \psi_{a_1 b_2 b_3} )\; ,
\end{split}
\end{align}
Here, terms inside a parenthesis have all their spinorial indices mutually contracted, see \eqref{eq:parenthesis} in appendix~\ref{app:invariants}. The matrices $\Gamma^X$, with $X=S,V,P$, are defined as 
\be
\Gamma^S=\one, \;\;\; \Gamma^V=\g^\m,
\;\;\; \Gamma^P=\g_5 \;. 
\ee
The letters $S$, $V$, and $P$ stand for scalar, vector, and pseudoscalar, respectively, in reference to the transformation properties of the associated bilinears under the rotation group, see appendix~\ref{app:invariants}.

The complex nature of the fields (i.e., the fact that an index in $\psi$ always needs to be contracted with an index in $\psib$ in order to form invariants) implies that the interaction graphs are bipartite; this means that the vertices can be divided in two sets, representing the $\psi$ and $\psib$ fields, respectively, in such a way that two vertices within the same set are never adjacent. One can immediately 
verifies that the first two graphs in figure~\ref{fig:vertices} are bipartite, while the third one is not.

The interaction \eqref{eq:vertex1} is built upon one of three possible invariants. We label them by $I_1^{X,\ell}$ where $X=S,V,P$ denotes the spinor structure~$\Gamma^X$, and $\ell=1,2,3$ describes the tensor structure. The interactions $I_1^{X,\ell}$ have the tensor index in the $\ell$-th position contracted between different bilinears. In other words, $I_1^{X,2}$ and $I_1^{X,3}$ are obtained as the two cyclic permutations of the color 
indices in the central picture of figure \ref{fig:vertices} (the color being a code name for the index location associated to an edge).
It is not difficult to verify that $I_0^X$ and $I_1^{X,\ell}$ are the only possible flavor-invariant interactions compatible with the requirements of renormalizability and Euclidean invariance.
The only non-trivial step is to show that the spinor contraction for the $I_1^{X,\ell}$ interactions can always be chosen as in figure \ref{fig:vertices}, rather than for example parallel to the color-1 edge in that picture. This is done by using standard Fierz identities or, in other words, the completeness relation~\eqref{eq:compl} of appendix~\ref{app:conventions}.

One can recognize that the $I_0^S$ interaction is nothing but the usual GN interaction in disguise, and thus it is invariant under a larger symmetry group, namely $U(N^3)$.%
\footnote{More general GN models, with also the $V$ and $P$ interactions have been studied in  \cite{Mitter:1974cy,Klimenko:1985ss,Bondi:1989nq}.}
In fact the kinetic term  has the same symmetry as well: it is the $I_1$ interactions that break such symmetry down to~$U(N)^3$.

In order to reduce the number of invariants, we can demand that the action be invariant under simultaneous and identical permutations of all the tensor indices, known also as {\it color symmetry}.
Such symmetry requires repackaging the $I_1^{X,\ell}$ interactions as
\be
I_1^X = \sum_{\ell=1}^3 I_1^{X,\ell}\;.
\ee
In this paper we will consider both models with and without color symmetry.\footnote{Color symmetric models are the most commonly studied in zero dimensions \cite{Gurau:2011xp,Gurau:book} 
but non-symmetric ones have also been considered, e.g. in  \cite{Benedetti:2015ara}.}

The most general renormalizable interacting action compatible with Euclidean, chiral, $U(N)^3$ and color symmetries thus contains six independent couplings, and reads\footnote{The minus sign is for later convenience. We recall that at finite $N$, and finite UV cutoff, the fermionic functional integrals are analytic around the origin. The good sign, if any, of the couplings can only be determined by the large-$N$ analysis of the renormalized theory.}
\be \label{eq:general-int-D}
S_{\rm int} = - \sum_{j=0}^1\ \sum_{X=S,V,P}\,  g_j^X\, \int d^2 x \; I_j^X  \;.
\ee
%

\subsection{Majorana fermions}

In the case of real Majorana fermions we define the free part of the action as
\be \label{eq:free-M}
S_{\rm free} =  \f12 \int d^2 x \; \psib_{abc} \slashed{\p} \psi_{abc}  \;,
\ee
where the different normalization is chosen for later convenience, taking into account that the spinors are real valued.

In constructing the interacting part of the action one encounters a new invariant, besides those we discussed for the Dirac case above. This is because here graphs representing the interaction vertex no longer need to be bipartite---recall that $\psib=\psi^{\rm t} \g_5$, and therefore $\psib$ transforms like $\psi$ with $U_{(i)}\in O(N)$, see \eqref{eq:flavor-sym}.
The new graph is tetrahedral (complete, four vertex), depicted on the right in figure \ref{fig:vertices},
\be
\label{eq:vertex2}
\begin{split}
I_2^{X,1} = &\big(\psib_{a_1 a_2 a_3} \Gamma^X \psi_{a_1 b_2 b_3}\big)\, \big( \psib_{b_1 a_2 b_3} \Gamma^X \psi_{b_1 b_2 a_3} \big)
\; ,
\end{split}
\ee
and generalizes the Klebanov-Tarnopolsky-Carrozza-Tanasa interaction to two dimensions \cite{Carrozza:2015adg,KlebanovTarnopolsky}.

In order to reduce the number of possible interactions, we would be tempted to require again color symmetry.
However, we must be a bit careful with how we average over colors for the $I_2^{X,\ell}$ interactions, because the symmetric structure of the complete graph associated to them implies crucial cancellations.
It turns out that there is no $I_2$ interaction invariant under the full color permutation group; there is however (precisely) one invariant if we only require symmetry under the alternating subgroup of the color permutation group, as we will show.

Let us introduce two sets of $I_2$ interactions, labelled by a sign $\pm$. The $I_{2,+}$ interactions are defined as
\begin{align}
I_{2,+}^{X,1} &= \big(\psib_{a_1 a_2 a_3} \Gamma^X \psi_{a_1 b_2 b_3}\big)\,\big( \psib_{b_1 a_2 b_3} \Gamma^X \psi_{b_1 b_2 a_3} \big)\;,\crcr
I_{2,+}^{X,2} &= \big(\psib_{a_1 a_2 a_3} \Gamma^X \psi_{b_1 a_2 b_3}\big)\,\big( \psib_{b_1 b_2 a_3} \Gamma^X \psi_{a_1 b_2 b_3} \big)\;, \\
I_{2,+}^{X,3} &= \big(\psib_{a_1 a_2 a_3} \Gamma^X \psi_{b_1 b_2 a_3}\big)\,\big( \psib_{a_1 b_2 b_3} \Gamma^X \psi_{b_1 a_2 b_3} \big)\;, \nonumber
\end{align} 
and the $I_{2,-}$ interactions as
\begin{align}
I_{2,-}^{X,1} &= \big(\psib_{a_1 a_2 a_3} \Gamma^X \psi_{a_1 b_2 b_3}\big)\,\big( \psib_{b_1 b_2 a_3} \Gamma^X \psi_{b_1 a_2 b_3} \big)\;,\crcr
I_{2,-}^{X,2} &= \big(\psib_{a_1 a_2 a_3} \Gamma^X \psi_{b_1 a_2 b_3}\big)\,\big( \psib_{a_1 b_2 b_3} \Gamma^X \psi_{b_1 b_2 a_3} \big) \;,\\
I_{2,-}^{X,3} &= \big(\psib_{a_1 a_2 a_3} \Gamma^X \psi_{b_1 b_2 a_3}\big)\,\big( \psib_{b_1 a_2 b_3} \Gamma^X \psi_{a_1 b_2 b_3} \big)\;.  \nonumber
\end{align} 
These definitions have been chosen so that, for any $X\in\{S, V , P\}$, $\ell \in \{1,2,3 \}$ and $i\in\{+,-\}$:
\begin{itemize}
\item the orbit of $I_{2,i}^{X,\ell}$ under the symmetric group $\cS_3$ (acting simultaneously on the three indices of all four tensors) is $\{ I_{2,j}^{X, \ell'}\ \vert\ j = \pm\,, \;\ell' = 1 , 2 , 3\}$;
\item the orbit of $I_{2,i}^{X,\ell}$ under the alternating group $\mathcal{A}_3 \subset \cS_3$ is $\{ I_{2,i}^{X, \ell'}\ \vert\ \ell' = 1 , 2 , 3\}$.
\end{itemize}

\

Now, the interactions $I_{2,+}^{X,\ell}$ and $I_{2,-}^{X,\ell}$ are not independent. Indeed, for Majorana spinors one has that $(\psib_{\mathbf{a}} \psi_{\mathbf{b}}) = (\psib_{\mathbf{b}} \psi_{\mathbf{a}})$, $(\psib_{\mathbf{a}} \gamma^\mu \psi_{\mathbf{b}}) = -(\psib_{\mathbf{b}} \gamma^\mu \psi_{\mathbf{a}})$ and $(\psib_{\mathbf{a}} \gamma_5 \psi_{\mathbf{b}}) = -(\psib_{\mathbf{b}} \gamma_5 \psi_{\mathbf{a}})$, which implies:
\begin{equation}
I_{2, +}^{S, \ell} = I_{2, -}^{S, \ell} \;,\qquad
I_{2, +}^{V, \ell} = - I_{2, -}^{V, \ell} \;,\qquad	
I_{2, +}^{P, \ell} = - I_{2, -}^{P, \ell}\;.
\end{equation}
We therefore conclude that the fully color-invariant $P$ and $V$ sectors are trivial. It would look like we have one invariant in the $S$ sector, i.e. $2\sum_{\ell} I_{2,+}^{S,\ell}$, but we will now show that this is also identically zero.

Let us determine the $\mathcal{A}_3$-invariant interactions, which we choose to parametrize in terms of the vertices $I_{2,+}$. The completeness relation \eqref{eq:compl} can be used to prove that\footnote{The addition operation on color labels is to be understood modulo $3$.}
\be
I_{2,+}^{S,\ell} = - \frac{1}{2} \sum_{X=S,V,P} I_{2,-}^{X, \ell - 1} = - \frac{1}{2} I_{2,+}^{S, \ell - 1} + \frac{1}{2} I_{2,+}^{V, \ell - 1} + \frac{1}{2} I_{2,+}^{P, \ell - 1}\,,
\ee
for any color $\ell$. By summing over $\ell$, this immediately implies:
\be \label{eq:I2SPV}
3 I_2^S = I_2^P + I_2^V\,,
\ee
where we have defined $I_2^X:= \sum_{\ell} I_{2,+}^{X,\ell}$. Therefore the $\cA_3$-invariant sector contains at most two independent coupling constants. 
Performing an odd permutation on the color indices in \eqref{eq:I2SPV} we find that the left-hand side is invariant, while the right-hand side picks a minus sign; therefore, both sides of the equation are zero. We are thus left with no invariants under $\cS_3$ and at most one invariant under $\cA_3$, which we could choose for example to be $I_2^P$. We are now going to show that this is in fact a non-trivial invariant, and that we can choose to write it in such a way that we do not have to perform a sum over colors.

To go further, we can invoke the completeness relations \eqref{eq:compl-2} and \eqref{eq:compl-3}
 to find:
\begin{align}
I_{2,+}^{V,\ell} &= - I_{2,-}^{S, \ell - 1} + I_{2,-}^{P, \ell - 1} = - I_{2,+}^{S, \ell - 1}  -  I_{2,+}^{P, \ell - 1}\,, \\
I_{2,+}^{P,\ell} &= -\frac{1}{2} I_{2,-}^{S, \ell - 1} + \frac{1}{2} I_{2,-}^{V, \ell - 1} - \frac{1}{2} I_{2,-}^{P, \ell - 1} = - \frac{1}{2} I_{2,+}^{S, \ell - 1}  - \frac{1}{2} I_{2,+}^{V, \ell - 1} + \frac{1}{2} I_{2,+}^{P, \ell - 1} \,.
\end{align}
From now on, let us drop the $\pm$ index and simply denote $I_{2,+}^{X,\ell}$ by $I_{2}^{X,\ell}$. For any given $\ell$, the Fierz relations show that we are free to parametrize our action with the basis $(I_{2}^{S,\ell}, I_{2}^{V,\ell}, I_{2}^{P,\ell})$. Let us denote by $\mathbf{g}_2^{\ell} = (g_2^{S, \ell}, g_2^{V, \ell}, g_2^{P, \ell})^t$ the coordinate vector in this basis, i.e.\ the interaction action in the sector $2$ is $S_{\mathrm{int},2} = - \sum_{X} g_2^{X,\ell} I_{2}^{X,\ell}$. The Fierz identities can be summarized in matrix form~by
\be 
\mathbf{g}_2^{\ell - 1} = F \mathbf{g}_2^{\ell}\,, \qquad F := \begin{pmatrix}  -1/2 & -1 & -1/2 \\
1/2 & 0 & -1/2 \\
1/2 & - 1 & 1/2\end{pmatrix} \,.
\ee
We can furthermore introduce the $\mathcal{A}_3$ averaging operator
\be 
A := \frac{1}{3} \left( \one + F + F^2 \right) = \begin{pmatrix}  0 & 0 & 0 \\
0 & 1/3 & -1/3 \\
0 & - 2/3 & 2/3\end{pmatrix} \,.
\ee
This matrix turns out to have a two-dimensional null eigenspace generated by $(1,0,0)^t$ and $(0,1,1)^t$, and a one-dimensional eigenspace with eigenvalue $1$ generated by $(0,-1,2)^t$. We recover in particular that two of the a priori allowed invariants identically vanish:
\be 
I_2^S = 0 = I_2^V + I_2^P\,.
\ee
The $\mathcal{A}_3$-invariant space of $I_2$ interactions is therefore one-dimensional and generated by the $\ell$-independent quantity (one can indeed check that $(0,-1,2)^t$ is an eigenvector of $F$ with eigenvalue~$1$):
\be \label{A3_inv}
I_2 := - I_{2}^{V, \ell} + 2 I_{2}^{P, \ell} \,.
\ee
Summing over $\ell$ we find $I_2=\f13(-I_2^V+2 I_2^P)=I_2^P$.

\bigskip

In summary, we conclude that: 1) there is no $I_2$ interaction invariant under the full permutation group $\cS_3$; 2) the space of $I_2$ interactions invariant under the subgroup of even permutations $\mathcal{A}_3$ is one-dimensional, and generated by \eqref{A3_inv}.

Remembering also that $I_0^P=I_0^V=0$ for Majorana fermions, we write the $\cA_3$-invariant interacting action for the Majorana case as
\be \label{eq:general-int-M}
S_{\rm int} = - \f14 \left(g_0 \int d^2 x \; I_0^S+ \sum_{X=S,V,P}  g_1^X \int d^2 x \; I_1^X +g_2 \int d^2 x \; I_2  \right)\;.
\ee
%

\section{Large $N$: Schwinger-Dyson equation and mass generation}
\label{sec:sde}

A powerful tool in the study of the large-$N$ limit is provided by the Schwinger-Dyson (SD) equations. In zero dimensions they typically allow to solve tensor models and find their critical behavior.
In one dimension, they play a crucial role for the SYK model and its tensorial generalizations, as they allow to identify a conformal IR regime and to solve the theory in that limit.
In this section we will study the SD equations for our Dirac and Majorana two-dimensional tensorial field theories, and in particular we will address the question of dynamical mass generation.

\subsection{Dirac case}
\label{sec:SDE-D}

In two dimensions we do not expect any breaking of continuous global symmetries. This fact is known  in the quantum field theory literature as Coleman theorem~\cite{Coleman:1973ci} and in the statistical mechanics literature as Mermin-Wagner theorem~\cite{Mermin:1966fe}, and we will come back to it in section~\ref{sec:Dirac-1color}.
Therefore, we should be allowed to assume $U(N)^3$-invariance of the theory, which implies that the propagator is proportional to the identity in tensor space:
\be
\langle \psi_{a_1 a_2 a_3}(x) \psib_{b_1 b_2 b_3}(x') \rangle = G(x,x')\, \d_{a_1 b_1} \d_{a_2 b_2} \d_{a_3 b_3} \;.
\ee
%
%
\begin{figure}[ht]
 \centering
             \includegraphics[scale = 0.75]{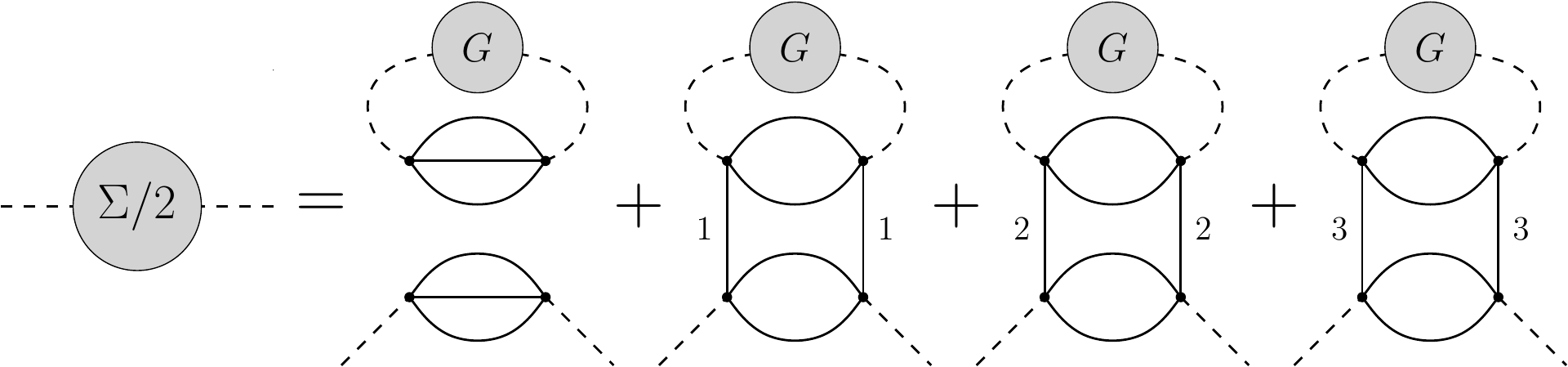}
  \caption{Graphical representation of the Schwinger-Dyson equation \eqref{eq:SD-Sigma-D} for the self-energy, at leading order in $N$.  For simplicity, we have not represented the contractions of spinor indices, and we have left the summation over $X$ implicit.}
\label{fig:sde}
\end{figure}

It is convenient to introduce a set of rescaled couplings defined as
\be
 \lambda_0^X = N^3 g_0^X\;, \;\;\; \lambda_1^X = N^2 g_1^X \;.
\ee
Then, the large-$N$ SD equations are:
\be
G^{-1}(x, x') = G_0^{-1}(x,x') - \Sigma (x,x') 
 \;,
\ee
\be \label{eq:SD-Sigma-D}
\Sigma (x,x') = -2 \sum_{\substack{X=S,V,P }}  (\lambda_0^X + 3 \lambda_1^X) \Tr[ G(x,x) \G^X ] \delta (x,x')\, \G^X 
\;,
\ee
where the factor 3 originates from the sum over colors in the $I_1$ interaction, see figure \ref{fig:sde}.
Combining the two, in Fourier space we find
\be \label{eq:SD-D}
\hat{G}^{-1}(p) = \im \slashed{p} + 2 \sum_{\substack{ X=S,V,P }}  (\lambda_0^X+3\lambda_1^X) \int \frac{d^2 q}{(2 \pi)^2}  \Tr[\hat{G}(q)  \G^X ]  \G^X  \;.
\ee
Notice that this has the structure $\hat{G}^{-1}(p) =\im \slashed{p} - \Sigma_0$, where $\Sigma_0$ is a momentum-independent quantity (and a matrix in spinor space). Furthermore, the two-point function should be invariant under the rotation group (including parity transformations), so that $\Sigma_0$ cannot be proportional to $\g^\m$ or $\g_5$. Therefore, $\Sigma_0$ must be proportional to the identity matrix, $\Sigma_0 = -m \one$; in the following we will omit writing explicitly the identity~$\one$ unless we need to highlight its presence.
This is consistent with \eqref{eq:SD-D} because $\g^\m$ and $\g_5$ are traceless, and it implies that the full propagator is $ (\im\slashed{p} + m)^{-1} = (-\im \slashed{p} + m ) / (p^2 + m^2 )$. Therefore, we can write a consistency equation for~$m$, also known as the \emph{gap equation}:
\be \label{eq:gap-D}
\begin{split}
m &= 2\left( \lambda_0^S + 3 \lambda_1^S \right) \int \frac{d^2 q}{(2 \pi)^2} \frac{\Tr[-\im\slashed{q} + m]}{q^2 + m^2} \\
&= 4 m  \left( \lambda_0^S + 3 \lambda_1^S \right) \int \frac{d^2 q}{(2 \pi)^2} \frac{1}{q^2 + m^2} \;.
\end{split}
\ee
Notice that the $P$ and $V$ interactions have dropped out of the equation.
Since the integral on the right-hand side is UV divergent, we introduce a UV cut-off $\Lambda$, and obtain:
\be
\int_{q^2 \leq \Lambda^2} \frac{d^2 q}{(2 \pi)^2} \frac{1}{q^2 + m^2} = \frac{1}{4 \pi} \ln( 1 + \frac{\Lambda^2}{m^2} ) \approx \frac{1}{4 \pi} \ln \frac{\Lambda^2}{m^2} \;.
\ee
The gap equation thus admits two solutions: $m=0$ and
\be \label{eq:dyn-mass-01}
m^2 = \Lambda^2 \exp\left( - \frac{ \pi}{ \lambda_0^S + 3 \lambda_1^S }\right)  \;.
\ee
By analogy with the GN model, we expect $m=0$ to correspond to an unstable vacuum, while \eqref{eq:dyn-mass-01} should be the physical vacuum. We will verify this expectation in the following section, by studying the effective potential in the intermediate field formalism.

\begin{figure}[t]
 \centering
        \begin{minipage}{0.45\textwidth}
            \centering 
             \includegraphics[scale = 0.7]{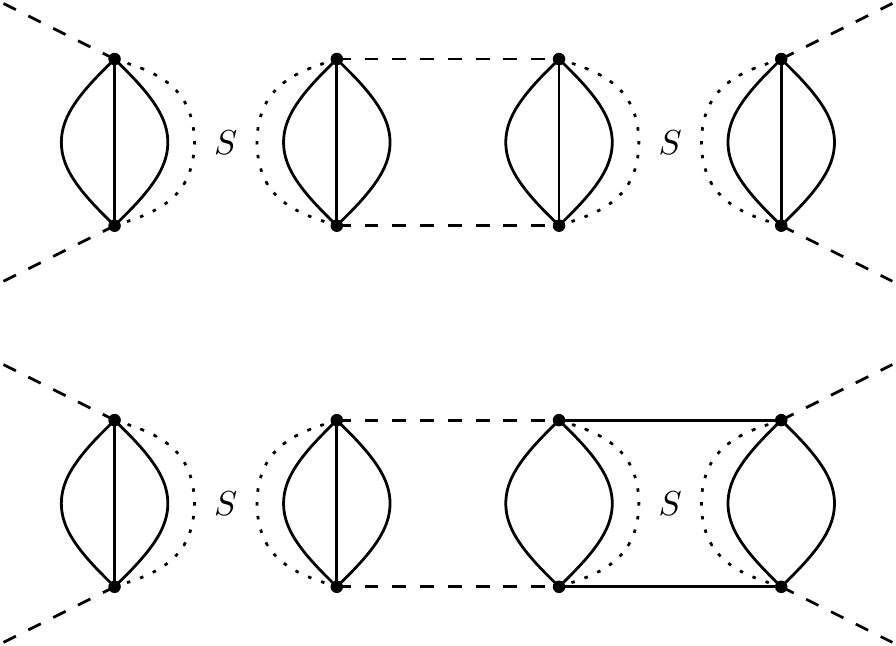}
        \end{minipage}
        \hspace{0.01\textwidth}
        \begin{minipage}{0.45\textwidth}
            \centering
            \includegraphics[scale = 0.7]{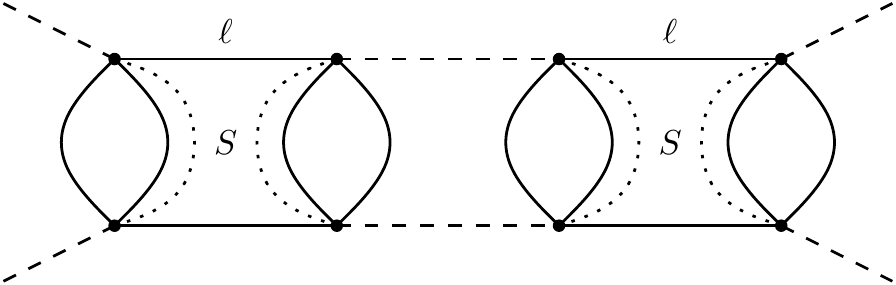}\\
            \vspace{.5cm}
            \includegraphics[scale = 0.7]{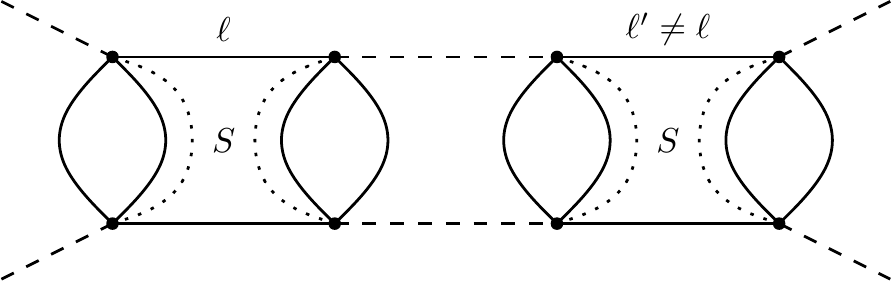}\\
        \end{minipage}
  \caption{\small{One-loop Feynman diagrams contributing to $\beta_1^S$ (top right) and $\beta_0^S$ (all the others). Here, the dashed lines represent free propagators.}}
\label{fig:1loop_beta02}
\end{figure}

\begin{figure}[ht]
 \centering
             \includegraphics[scale = 0.85]{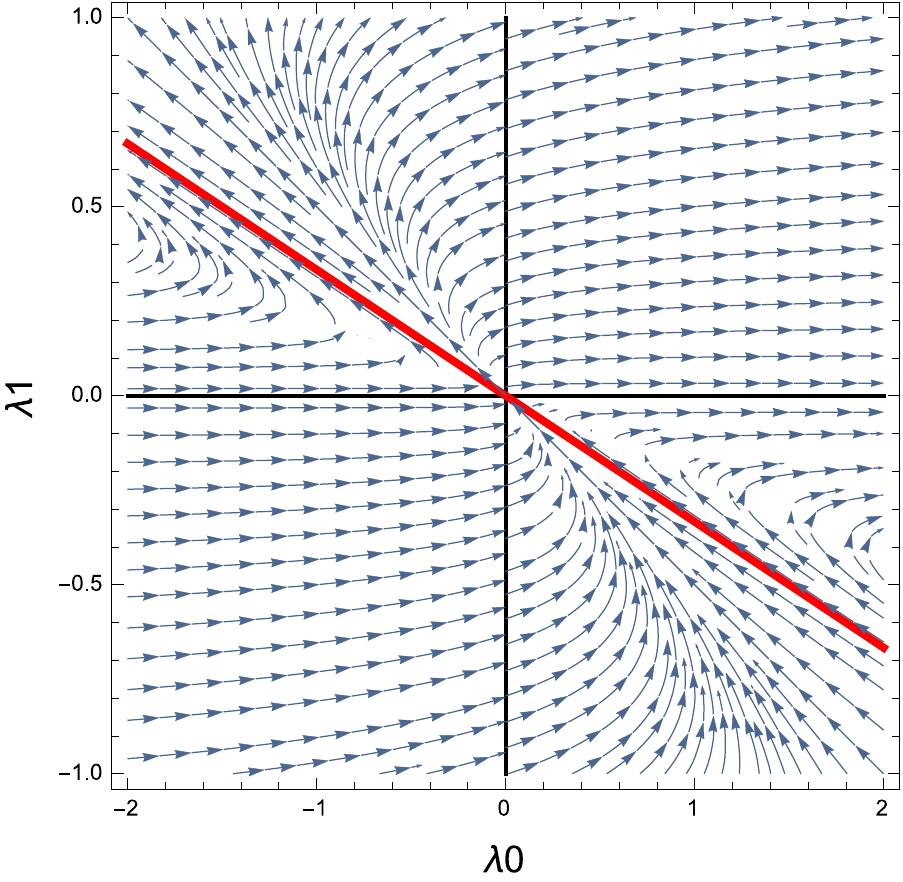}
  \caption{Flow of the beta functions \eqref{eq:beta1} and \eqref{eq:beta0}. Arrows point towards the IR. The red line is the locus $\l_0+3\l_1=0$.}
\label{fig:betaPlot}
\end{figure}

The non-zero solution \eqref{eq:dyn-mass-01} is a non-perturbatively generated mass, which breaks the discrete chiral invariance of the bare theory.
Requiring that this mass is invariant under a change of UV scale $\Lambda$ yields the renormalization group equation:
\be
0 = \left( \Lambda \frac{\partial }{\partial \Lambda} + \sum_{\substack{ i= 0, 1 \\ X=S,V,P }}  \beta_i^X \frac{\partial }{\partial \lambda_i^X}  \right) m^2 = \left( 2 + \frac{\pi}{(  \lambda_0^S + 3 \lambda_1^S  )^2 } \left( \beta_0^S + 3\beta_1^S \right)   \right) m^2\,,
\ee
where $\beta_i^X := \Lambda \partial \lambda_i^X / \partial \Lambda$. Hence we must have:
\be\label{eq:2beta}
\beta_0^S + 3\beta_1^S  = -\frac{2}{ \pi}( \lambda_0^S + 3\lambda_1^S )^2 \;.
\ee
Moreover, by direct examination of the possible Feynman diagrams contributing to the two flows (see figure \ref{fig:1loop_beta02}), one notices that among the  terms in $(\lambda_1^S)^2$, only the top right graph contribute to $\beta_1^S$, while all the others contribute to $\beta_0^S$. Hence we may conclude that the single beta functions are:
\begin{align} \label{eq:beta1}
\beta_1^S &= - \frac{2(\lambda_1^S)^2}{\pi} \,, \\ 
\label{eq:beta0}
\beta_0^S &= - \frac{2}{\pi} \Big(( \lambda_0^S)^2 + 6 \lambda_0^S \lambda_1^S + 6 ( \lambda_1^S)^2 \Big) \,.
\end{align}
Their associated flow is depicted in figure \ref{fig:betaPlot}.

In conclusion, we have the same situation as in the standard GN model: asymptotic freedom (for both $\lambda_0^S$ and $\lambda_1^S$; we have not discussed the beta functions for the $P$ and $V$ couplings as they do not affect the 2-point function) and mass generation. Notice that \eqref{eq:beta0} leads to \eqref{eq:GN-fullBeta} for $\l^S_1=0$ and $\l^S_0=\l$, as expected.

More interestingly, we see from \eqref{eq:dyn-mass-01} that we now have the opportunity to impose the massless condition: 
\be\label{massless_cond}
3 \lambda_1^S = - \lambda_0^S > 0\,.
\ee
Indeed, this subspace of theories is preserved by the RG flow \eqref{eq:2beta} (as seen also in  figure \ref{fig:betaPlot}), and the signs of the beta functions are consistent with asymptotic freedom:
\be
\beta_1^S < 0 \qquad \mathrm{and} \qquad \beta_0^S = + \frac{(\lambda_0^S)^2}{\pi} > 0 \,,
\ee
in such a way that
\be
3 \lambda_1^S ( \Lambda ) = - \lambda_0^S (\Lambda ) = \frac{g}{1 + \frac{g}{\pi} \ln( \Lambda / \Lambda_0)}\,,
\ee
where $g>0$ is the value of $3\lambda_1^S$ at some reference scale $\Lambda_0$.
However, in view of equation \eqref{eq:2beta}, the massless condition \eqref{massless_cond} is IR unstable, and therefore it cannot be reached by a generic theory. Nevertheless, given that the condition  $ \lambda_0^S + 3\lambda_1^S=0$ is RG invariant, one could define the theory to live in such a subspace. In this case we would seem to have a massless, although non-conformal theory: in fact there is still a non-trivial renormalization group running (defining a flow along the red curve in figure \ref{fig:betaPlot}), hence the theory is not scale invariant. The only zero of the beta functions is at the origin, i.e. the free theory.

It is however premature to draw any conclusion on the existence of a massless theory. There are at least two possible pitfalls: the condition $ \lambda_0^S + 3\lambda_1^S=0$ could correspond to a trivial (non-interacting) theory in disguise; or it could correspond to a range of parameters for which the theory is unstable.
In particular it is natural to worry about the second possibility, given that the massless condition forces the two couplings to have opposite sign.
In order to check whether any such negative scenario is realized or not, we could look at the two-point function for the composite operator $\psib_{\bf a} \psi_{\bf a}$ and check whether it decomposes trivially or whether it shows any tachyonic poles.
An easier method to effectively do such computation is to introduce an intermediate field representation, as we will do in section \ref{sec:Dirac-1color}.

\subsection{Majorana case}
\label{sec:SDE-M}

In the case of real Majorana fermions we have a minor and a major change with respect to the complex Dirac case. The minor change is some combinatorial factors. The major change is the presence of $I_2$ in the action \eqref{eq:general-int-M}.

\begin{figure}[ht]
        \begin{minipage}{0.4\textwidth}
            \centering 
             \includegraphics[scale = 0.75]{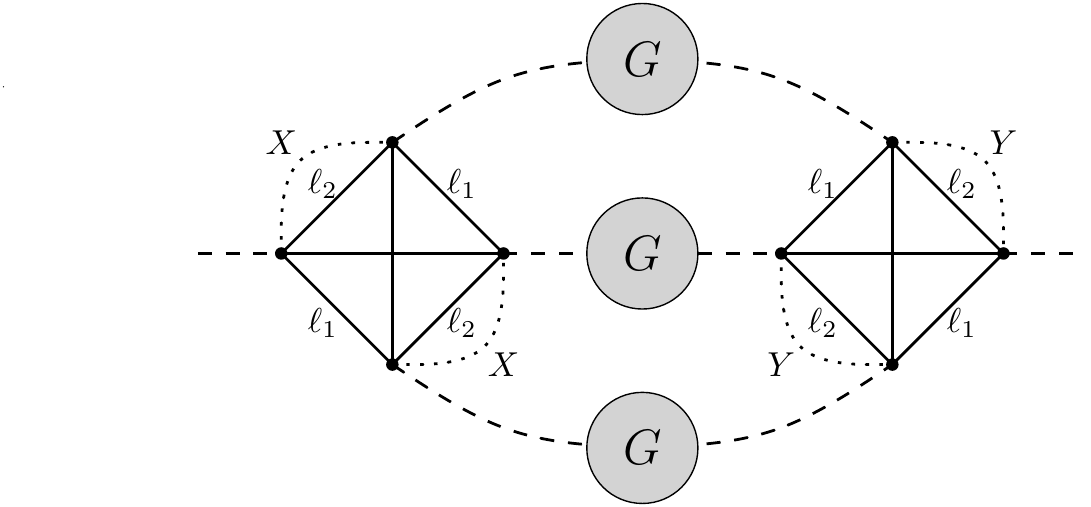}
        \end{minipage}
        \hspace{0.05\textwidth}
        \begin{minipage}{0.4\textwidth}
            \centering
            \includegraphics[scale = 0.75]{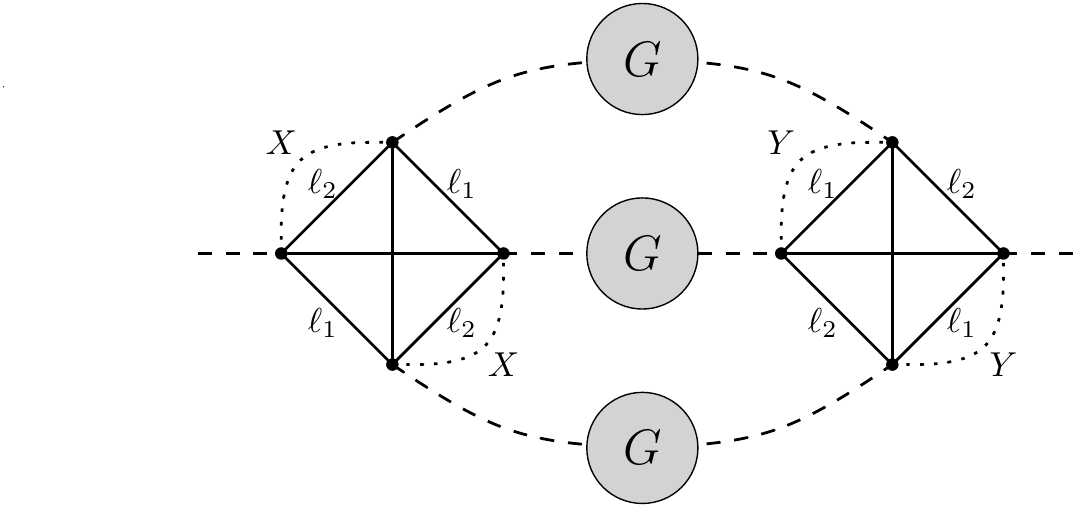}
        \end{minipage}
 \caption{\small{Leading order contributions to the self-energy resulting from interactions of type $I_2$: the graph on the left involves a trace over the spinorial indices, while the graph on the right does not.}}
\label{fig:sde_I2}
\end{figure}

The large-$N$ limit for real rank-3 tensors with $O(N)^3$ invariance has been studied in \cite{Carrozza:2015adg}, but here we have two main differences: the spacetime dependence, and the spinor indices. The latter lead to two types of contributions from the $I_2$ vertices: one without and one with a trace on spinorial indices, both depicted in figure \ref{fig:sde_I2}.

Defining the rescaled couplings as (in agreement with \cite{Carrozza:2015adg})
\be
 \lambda_0 = N^3 g_0\;, \;\;\; \lambda_1^X = N^2 g_1^X ;, \;\;\; \lambda_2 = N^{3/2} g_2 \;,
\ee
 the  large-$N$ SD equations are:
\be
G^{-1}(x, x') = G_0^{-1}(x,x') - \Sigma (x,x')
 \;,
\ee
and\footnote{Note that the last two terms have the same sign, whereas one might think that they should differ in sign due to the fermionic trace in the last term. The reason is  the antisymmetry of $(\psib_{\mathbf{a}} \gamma_5 \psi_{\mathbf{b}})$: carefully keeping track of the $\psi$'s and $\psib$'s along the chain of fermions, one notices that in the case on the right panel of figure \ref{fig:sde_I2} exactly one pair $(\psib_{\mathbf{a}} \gamma_5 \psi_{\mathbf{b}})$ must be transposed, yielding a global minus sign.}
\be \label{eq:Sigma-M}
\begin{split}
\Sigma (x,x') = & - \sum_{\substack{ X=S,V,P }}  (\lambda_0^X+3\lambda_1^X) \Tr[ G(x,x) \G^X ] \delta (x,x')  \G^X \\
& - 3 (\lambda_2)^2 \Big( 2\,  \g_5 G (x,x') \g_5 G(x',x) \g_5 G(x,x') \g_5  \\
&\qquad\qquad\  + \g_5 G(x,x') \g_5  \Tr[G(x',x) \g_5 G(x,x') \g_5 ] \Big)
\;,
\end{split}
\ee
where $\l_0^S=\l_0$, and $\l_0^V=\l_0^P=0$.
In momentum space we have
\be \label{eq:SDE-p}
\hat{G}(p)^{-1} = \im \slashed{p} - \hat{\Sigma} (p)  \;,
\ee
and
\be \label{eq:Sigma_p-M}
\begin{split}
\hat{\Sigma}(p) = & - \sum_{\substack{ X=S,V,P }}  (\lambda_0^X+3\lambda_1^X) \int \frac{d^2 q}{(2 \pi)^2}  \Tr[\hat{G}(q)  \G^X ]  \G^X  \\
& - 3 (\lambda_2)^2 \int \frac{d^2 q_1 d^2 q_2 d^2 q_3}{(2 \pi)^4} \delta( p - \sum_i q_i) \\
&\qquad\qquad  \times \hat{G}(-q_3) \left( 2 \hat{G}(-q_2) \hat{G}(-q_1) + \Tr\left[ \hat{G}(-q_2) \hat{G}(-q_1) \right] \right)
\;.
\end{split}
\ee

\begin{figure}[ht]
 \centering
             \includegraphics[scale = 0.75]{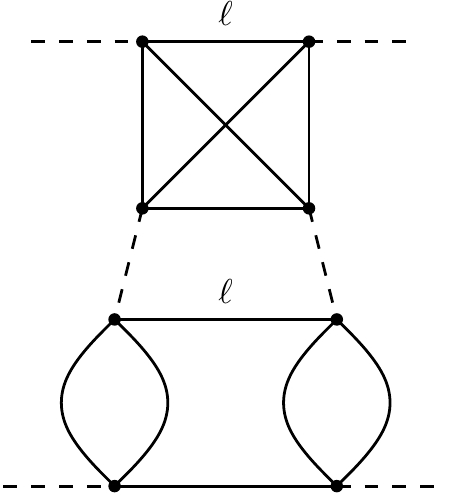}
  \caption{Structure of the leading contributions to the one-loop beta function of $\lambda_2$: such diagrams are suppressed by a factor of $1/N$ with respect to the natural scaling of $\lambda_2$.}
\label{fig:beta2}
\end{figure}

Unlike the tadpole contributions proportional to $\l_0^S$ and $\l_1^S$, the sunset ones proportional to $\l_2$ have a non-trivial dependence on the external momentum $p$, and therefore we no longer can assume a simple momentum dependence for the 2-point function.
Assuming no parity breaking we can write the Fourier-transformed inverse 2-point function as
\be \label{eq:G-ansatz}
\hat{G}(p)^{-1} = A(p^2) \im \slashed{p} + B(p^2) \;.
\ee
Locality imposes that  $A(p^2)$ and $B(p^2)$ have no singularities for real $p$, hence $\hat{G}(0)^{-1} =B(0)$. Assuming also that $A^2(p^2) p^2 + B^2(p^2)$ is a real and monotonically increasing function of $p$, we see that we will avoid tachyonic poles in the 2-point function if $B^2(0)\geq 0$, and that for $B(0)=0$ we have a massless theory.
Unfortunately, solving the SD equations for $A(p^2)$ and $B(p^2)$ is beyond~reach.
In fact our goal is more modest, as we would like to understand if a conformal theory is a possible solution of the SD equations for some values of the couplings (i.e. if there is a non-trivial fixed point). To that end, we could in principle try to assume that the 2-point function takes the form 
\be \label{eq:conf-prop}
\hat G(p) = - \im b \frac{\slashed{p}}{p^{2\alpha}} \,,
\ee
where we introduced two parameters $b$ and $\alpha$ to be determined from the SD equation.
Na\"ive dimensional analysis of the equation then would fix $\a=1$, i.e. as a consequence of the large-$N$ SD equation a conformal theory necessarily has zero anomalous dimension.
However, besides being UV divergent, in the massless case the loop integrals are also IR divergent, and we are forced to reintroduce a mass in order to regularize them. 
In practice we will use the following recipe to search for non-trivial fixed points:
\begin{enumerate}
\item consider the crude approximation of the SD equations in which we take $b \hat{G}(p)^{-1} = \im\slashed{p} + m$;
\item write a gap equation for the mass and look for conditions leading to $m=0$, as done before;
\item impose that the conditions obtained on the couplings are stable under radiative corrections;
\item impose that the remaining coupling constants have zero beta function.
\end{enumerate}
We will find that step 2 imposes separate conditions on the couplings of type-2 interactions\footnote{Although we have up to here considered only color-symmetric interactions, and thus a single type-2 interaction, we will soon consider also non-color-symmetric variants, with multiple couplings.} and on those of the type-0 and type-1 interactions. For step 3 then it will suffice to require non-renormalization of $\l_0$ and $\l_1$ from $I_2$. In fact it turns out that at leading order in $1/N$ there is no renormalization of type-2 vertices, hence any homogeneous condition on $\l_2$ couplings will be preserved under renormalization. It is easy to see that this must hold at one-loop, since the leading contributions to the 4-point function have a colored structure as shown in figure \ref{fig:beta2}: such diagrams scale as $(1/N)^2\times(1/N)^{3/2}\times N = (1/N)^{5/2}$, and are therefore suppressed by a factor $1/N$ with respect to the natural scaling of $I_2$. Furthermore, the results of references \cite{Gurau:2016lzk} and \cite{Carrozza:2015adg} guarantee that such a suppression occurs at any order in the Feynman expansion. 
As a consequence of the last observation, the requirement that $\l_2$ couplings have zero beta function will amount to imposing no wave-function renormalization. See appendix \ref{app:beta2} for a derivation of the beta function of $\l_2$ at leading order in both $1/N$ and the coupling.

Let us now proceed to apply this recipe to the above SD equations.
We start by plugging into equation \eqref{eq:Sigma_p-M} the ansatz for $\hat{G}(p)$ \eqref{eq:G-ansatz}, further specifying $A(p^2)=1$, $B(p^2)=m$ and setting $p=0$. After a lengthy but straightforward computation we obtain the following gap equation
\be \label{eq:gap-M}
m =  2 m b^2 (\l_0^S +3\l_1^S) \cJ  
+ 12 m (\lambda_2)^2 b^4 \cJ^2 \;,
\ee
where $\cJ$ is the divergent integral encountered before, i.e.
\be \label{eq:J-int}
\cJ :=  \int \frac{d^2 q}{(2 \pi)^2}  \ \frac{1}{q^2 + m^2} \approx \frac{1}{2 \pi} \ln \frac{\Lambda}{m} \;.
\ee

As before, we still have a solution with $m=0$. We would again expect it to correspond to an unstable vacuum, but verifying this expectation is more complicated in the present case because we do not have a useful intermediate field representation of $I_2$.
In any case, if tachyonic poles depend continuously on $\l_2$, we would expect this vacuum to remain unstable, at least for small $\l_2$.
We also have two non-trivial solutions:
\be \label{eq:dyn-mass-M}
m_\pm = \L \exp\left(  \f{a_1\pm\sqrt{a_1^2 + 4 a_2}}{2a_2}  \right) \;, 
\ee
where
\begin{equation}
a_1 =  \f{b^2}{\pi} (\l_0^S +3\l_1^S) \;,\qquad
a_2 = \f{12 b^4}{(2\pi)^2}(\lambda_2)^2 \;.
\end{equation}
The solution \eqref{eq:dyn-mass-M} is non-zero whenever $\lambda_2 \neq 0$, which suggests that a non-perturbative mass is always generated in the $\cA_3$-invariant model, irrespectively of the values of the coupling constants. This implies that such a theory can never flow towards a conformal fixed point.
For $\l_0^S +3\l_1^S>0$, the physically relevant solution is given by $m_-$. In fact, in the limit $\l_2\to 0$ this yields back \eqref{eq:dyn-mass-01},
\be
\lim_{\l_2\to 0} m_- =  \L \exp\left( - \f{\pi}{b^2 (\l_0^S+3\l_1^S)}  \right) \;.
\ee
On the other hand, $m_+$ diverges in the same limit, which suggests that the theory in this case does not admit a weak-coupling limit.

One may wonder whether the conclusion of this analysis would still hold in a more general model without  color symmetry. To settle this question, let us determine the gap equation associated to the most generic action with only $I_2$ interactions:
\be\label{S2_general}
S_{\mathrm{int},2 }= \sum_{X=S,V,P} \lambda_2^{X,1} I_2^{X,1}\,,
\ee
where we work in the $\ell=1$ representation for definiteness. As we have determined in section \ref{sec:models}, this covers all possible type-$2$ interactions, with or without invariance under permutation of the colors. The parametrization \eqref{S2_general} has the computational advantage that all the Feynman diagrams it generates in the self-energy have a spinorial trace, as in the left panel of figure \ref{fig:sde_I2}. The self-energy now reads:
\begin{align} \nn
\Sigma(p) &= - \int \frac{d^2 q_1 d^2 q_2 d^2 q_3}{(2 \pi)^4} \delta( p - \sum_i q_i) \\ \nn
& \quad  \left[ \; (\lambda_2^{S,1})^2 \, \hat{G}(q_3) \Tr\left[ \hat{G}(-q_2)  \hat{G}(q_1) \right] \right. \\
& \quad + (\lambda_2^{P,1})^2 \, \hat{G}(-q_3) \Tr\left[ \hat{G}(-q_2) \hat{G}(-q_1) \right] \\ \nn
& \quad + (\lambda_2^{V,1})^2 \, \gamma_\mu \hat{G}(q_3) \gamma_\nu \Tr\left[ \hat{G}(-q_2) \gamma^\mu \hat{G}(q_1) \gamma^\nu \right] \\ \nn
&\left. \quad +  \lambda_2^{V,1} \lambda_2^{P,1} \left( \gamma_\mu \hat{G}(q_3) \gamma_5 \Tr\left[ \hat{G}(-q_2) \gamma^\mu \hat{G}(q_1) \gamma_5 \right] + \gamma_5 \hat{G}(q_3) \gamma_\mu \Tr\left[ \hat{G}(-q_2) \gamma_5 \hat{G}(q_1) \gamma^\mu \right] \right) \; \right]\;.
\end{align}
Note that there is no term in $\lambda_2^{V,1} \lambda_2^{S,1}$ or $\lambda_2^{S,1} \lambda_2^{P,1}$ as these cancel by antisymmetry in $q_1 \leftrightarrow q_2$. After a somewhat lengthy but straightforward calculation, one arrives at:
\be
\begin{split}
\frac{\Sigma(0)}{m b^3} 
&= - \left( 3 (\lambda_2^{S,1})^2 + (\lambda_2^{P,1})^2 + 4 (\lambda_2^{V,1})^2 + 4 \lambda_2^{V,1} \lambda_2^{P,1} \right) \cI \\
&\qquad - \left( - (\lambda_2^{S,1})^2 + (\lambda_2^{P,1})^2 - 4 \lambda_2^{V,1} \lambda_2^{P,1} \right) \cJ^2 \; ,
\end{split}
\ee
where $\cJ$ was defined in \eqref{eq:J-int}, while $\cI$ is an $m$-independent finite integral which is the subject of much literature (e.g. \cite{Bloch:2013tra,Adams:2014vja}), i.e.
\be
\cI := \int \frac{d^2 q d^2 p}{(2 \pi)^4} \frac{m^2}{ (p^2 + m^2 )(q^2 + m^2) [(p+q)^2 + m^2] } \;.
\ee
Equation \eqref{eq:gap-M} is recovered for $\lambda_2^{P,1} / 2 = - \lambda_2^{V,1} = \lambda_2$ and $\lambda_2^{S,1} = 0$, as it should. Interestingly, one realizes that the divergent term proportional to $\cJ^2$, which is the source of the spontaneous generation of mass by $I_2$ interactions, is cancelled in the gap equation whenever
\be\label{eq:I2_cond0}
- (\lambda_2^{S,1})^2 + (\lambda_2^{P,1})^2 - 4 \lambda_2^{V,1} \lambda_2^{P,1} = 0\,.
\ee

Following our recipe further, we should now impose that the condition  \eqref{eq:I2_cond0}, together with $\l_0^S=\l_1^S=0$, is stable under radiative corrections.
As explained before, since equation  \eqref{eq:I2_cond0} is homogeneous of degree two in the couplings of type-2 interactions, it is stable under radiative corrections at leading order in $1/N$, because the only possible source of flow of such couplings comes from the wave-function renormalization, which however is the same for all of them.
Therefore, it only remains to look for additional conditions which ensure that no $I_0^{S,\ell}$ and $I_1^{S,\ell}$  interactions are generated by the RG flow at leading order in $N$, so that our SD equation with only type-2 interactions remains consistent. It is easy to check that type-0 interactions being topologically disconnected, they are never generated by the connected type-2 interactions. 
For the type-1 interactions, at first sight this looks doomed, since  one would naively need to impose three conditions (one for each $\ell=1,2,3$, due to the color non-invariance) in a two-dimensional theory space (defined by \eqref{eq:I2_cond0}). As it turns out, we will see that these conditions are not independent and admit non-trivial solutions.

\begin{figure}[ht]
 \centering
        \begin{minipage}{0.25\textwidth}
            \centering 
             \includegraphics[scale = 0.8]{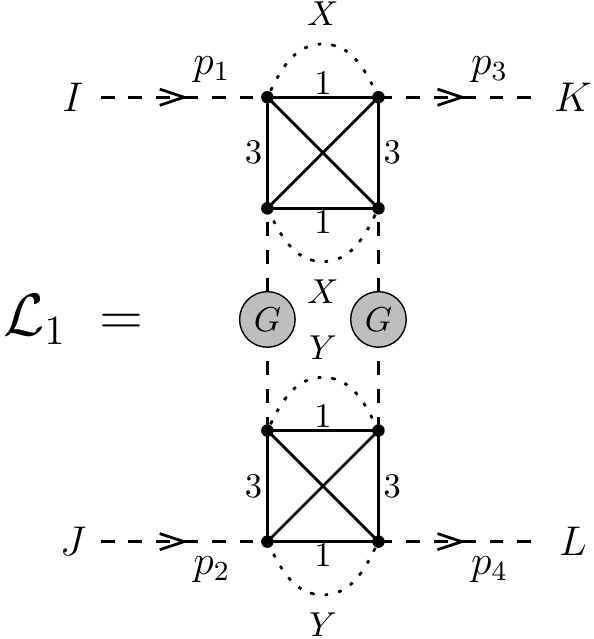}
        \end{minipage}
        \hspace{6mm}
        \begin{minipage}{0.25\textwidth}
            \centering
            \includegraphics[scale = 0.8]{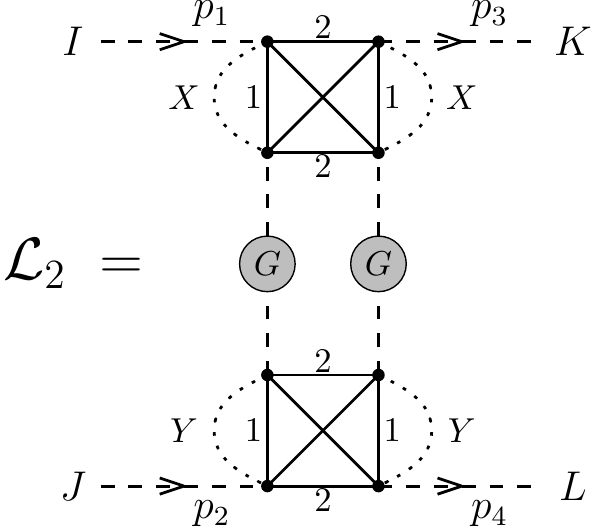}
        \end{minipage}
        \hspace{6mm}
        \begin{minipage}{0.25\textwidth}
            \centering
            \includegraphics[scale = 0.8]{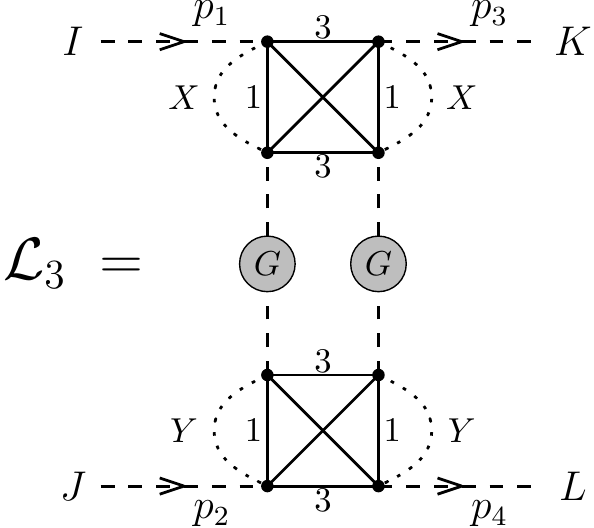}
        \end{minipage}
  \caption{The three types of ladders contributing to the leading-order 4-point function.}
\label{fig:ladders}
\end{figure}

The leading order 4-point functions of melonic theories are sums of ladder diagrams (see e.g.\ \cite{Gurau:2016lzk}). In the present theory, we have three types of elementary ladders, as shown in figure \ref{fig:ladders}. The effective $4$-point interactions generated by the ladders $\cL_\ell$ are of type $I_1^{X, \ell}$. Note that for $\ell =1$ one needs to use the Fierz identities to re-express the naturally generated interactions in our chosen basis. 

In the following, we will assume that $\hat{G}$ is of the form \eqref{eq:G-ansatz} with $B(p^2) =0$. The kernel associated to the diagram $\cL_3$ is (we keep the conservation of external momentum implicit)
\be
\lambda_2^{X,1} \lambda_2^{Y,1} \int \frac{d^2 q_1 d^2 q_2}{(2 \pi)^2} \delta( p_1 - p_3 - q_1 - q_2) \left[ \Gamma^X \hat{G} (q_1) \Gamma^Y \right]_{IJ} \left[ \Gamma^X \hat{G}(q_2) \Gamma^Y \right]_{KL}\;.
\ee
Forgetting for a moment about IR divergences, at $p_1 = p_2$, this takes the form 
\be
\frac{1}{2} \lambda_2^{X,1} \lambda_2^{Y,1} \int \frac{d^2 q}{(2 \pi)^2} \frac{1}{A^2(q^2) q^2} \left[ \Gamma^X \gamma^\mu \Gamma^Y \right]_{IJ} \left[ \Gamma^X \gamma_\mu \Gamma^Y \right]_{KL}\;.
\ee 
Most of these contributions are proportional to $[\gamma^\mu]_{IJ} [\gamma_\mu]_{KL}$ (resp. $[\gamma_5]_{IJ} [\gamma_5]_{KL}$) and therefore renormalize $I_{1}^{V, 3}$ (resp. $I_{1}^{P, 3}$). The only contributions which matter to our discussion are from $\{X,Y\}=\{ S,V \}$ and $\{ X,Y\}=\{V,P\}$. For the former we find an amplitude proportional to
\be
\left[ \gamma^\mu \gamma^\nu \right]_{IJ} \left[ \gamma_\mu \gamma_\nu \right]_{KL} = 2 \delta_{IJ} \delta_{KL} - 2 [\gamma_5]_{IJ} [\gamma_5]_{KL}\;,
\ee
while the latter is proportional to:
\be
\left[ \gamma^\nu \gamma^\mu \gamma_5 \right]_{IJ} \left[ \gamma_\nu \gamma_\mu \gamma_5 \right]_{KL} = - 2 \delta_{IJ} \delta_{KL} + 2 [\gamma_5]_{IJ} [\gamma_5]_{KL}\;.
\ee
In order to maintain $\lambda_{1}^{S,3} = 0$, the terms in $\delta_{IJ} \delta_{KL}$ must cancel each other, yielding the condition:
\be 
\lambda_2^{V,1} \lambda_2^{P,1} = \lambda_2^{V,1} \lambda_2^{S,1}\;.
\ee

We can proceed similarly for the ladder $\cL_2$, but by symmetry of the colors $2$ and $3$ in this problem one does not get any new condition: $\lambda_{1}^{S,2} = 0$ imposes no more than $\lambda_2^{V,1} \lambda_2^{P,1} = \lambda_2^{V,1} \lambda_2^{S,1}$.

There is one extra condition to be derived from $\lambda_{1}^{S,1} = 0$. The relevant ladder is $\cL_1$, with amplitude: 
\be
- \lambda_2^{X,1} \lambda_2^{Y,1} \int \frac{d^2 q_1 d^2 q_2}{(2 \pi)^2} \delta( p_1 - p_3 - q_1 - q_2) \left[ \G^X \right]_{IK} \left[ \G^Y \right]_{JL} \Tr \left[ \Gamma^X \hat{G} (-q_2) \Gamma^Y \hat{G}(q_1)\right]\;.
\ee
For diagrams with $\{X,Y\} \in \{ \{S,V\},  \{V, P\} \}$, the trace vanishes identically, while for $\{ X, Y\} = \{ S, P\}$ it yields an integrand which is antisymmetric in $q_1 \leftrightarrow q_2$ and therefore integrates to~$0$. From the identity $\Tr[\gamma^{\mu} \gamma^{\rho} \gamma_\mu \gamma^{\sigma}] = 0$, one infers that the $\{V, V\}$ contributions also vanish. We are left with the $\{ S, S\}$ and $\{ P, P\}$ diagrams, the first of which generates a contribution proportional to $\delta_{IK} \delta_{JL}$, while the second produces a term in $\left[ \gamma_5 \right]_{IK} \left[ \gamma_5 \right]_{JL}$. Furthermore, these two contributions come with opposite multiplicative constants (as a consequence of $\{ \gamma_\mu , \gamma_5 \} = 0$ and $\gamma_5^2 = 1$). By the completeness relations \eqref{eq:compl} and \eqref{eq:compl-2}, the corrections to $\lambda_2^{S,1}$ generated by these two terms compensate each other whenever
\be 
(\lambda_2^{S,1})^2 = (\lambda_2^{P,1})^2\;,
\ee
which is the last consistency condition we can derive from the 4-point function. 

In summary, we have learned from steps 2 and 3 of our recipe that we might obtain a massless theory provided one of the following two sets of conditions is satisfied:
\be\label{eq:I2_cond1}
(\lambda_2^{S,1})^2 = (\lambda_2^{P,1})^2 = \lambda^2 \neq 0 \qquad \mathrm{and} \qquad \lambda_2^{V,1} = 0\;,  
\ee
or 
\be\label{eq:I2_cond2}
(\lambda_2^{S,1})^2 = (\lambda_2^{P,1})^2 = 0 \qquad \mathrm{and} \qquad \lambda_2^{V,1} = \lambda \neq 0 \;.
\ee
Interestingly, such conditions are satisfied by models invariant under the continuous chiral symmetry \eqref{eq:cct1}.

Step 4 just amounts now to looking for possible conditions that avoid wave function renormalization. Unfortunately the previous steps left us with only one free coupling, thus either we are lucky and there is no need of wave function renormalization or such further requirement will lead us to a free theory.
Assuming that one of the two set of conditions \eqref{eq:I2_cond1} and \eqref{eq:I2_cond2} hold, the free-energy $\Sigma(p)$ evaluated at first order in the external momentum $p$ is:
\be
\begin{split}
\Sigma(p) &= 2 \im b^3 \left( (\lambda_2^{S,1})^2 + (\lambda_2^{P,1})^2 + 2 (\lambda_2^{V,1})^2 \right) \int \frac{d^2 q_1 d^2 q_2 d^2 q_3}{(2 \pi)^4} \delta( p - \sum_i q_i) \frac{(q_1 \cdot q_2) \slashed{q_3}}{q_1^2 q_2^2 q_3^2} \\
&=  4 \im b^3  \lambda^2  \slashed{p}   \int \frac{d^2 q_1 d^2 q_2 d^2 q_3}{(2 \pi)^4} \delta( p - \sum_i q_i) \frac{(q_1 \cdot q_2) \slashed{q_3}}{q_1^2 q_2^2 q_3^2}\,,
\end{split}
\ee
which is of course both IR and UV divergent. In dimensional regularization one finds a simple pole divergence (see appendix \ref{app:beta2}), or a logarithmic divergence in a momentum cutoff.
This means that we have a non-trivial wave function renormalization, which can only be eliminated by the condition $\l=0$, corresponding to a free theory.
In conclusion, the conditions \eqref{eq:I2_cond1} and \eqref{eq:I2_cond2} do not correspond to a fixed point.

In  appendix \ref{app:beta2} we show that the coupling $\l$ in equation \eqref{eq:I2_cond1} has, at leading order in $1/N$ and $\l$, the following beta function:
\be \label{eq:beta2}
\b_2 =  \f{3}{\pi^2} \l^3 \;.
\ee
As a consequence, unlike $\l_0^S$ and $\l_1^S$, the coupling is not asymptotically free, it is IR free for both signs of the coupling. Such a result derives from the fact that at leading order in $1/N$ the beta function starts at two-loop order. The sign in  \eqref{eq:beta2} is thus compatible with results from the usual GN model, where the leading terms in $1/N$ of the two loop contributions to the beta functions have the same positive sign as our \eqref{eq:beta2} (see for example \cite{Bondi:1989nq}). The key difference is that for the GN model the two-loop part of the beta function is subleading in $1/N$ with respect to the one-loop part, while here the situation is reversed.

We conclude this section by noticing that in $d=2-\eps$ dimensions the beta function becomes
\be
\b_2 = -\epsilon \l+ \f{3}{\pi^2} \l^3 \;,
\ee
thus showing a non-trivial IR fixed point of order $\sqrt{\epsilon}$, both at positive and negative coupling.
On the contrary, no UV fixed point is found in $d=2+\epsilon$ dimensions. The situation is thus reversed with respect to the GN model.
Such result provides support for the conjecture in  \cite{Prakash:2017hwq}  that the theory in $d=2-\eps$ dimensions flows to a weakly interacting IR fixed point for small $\epsilon$.\footnote{Although in appendix  \ref{app:beta2} we only discuss the models introduced in the present paper, it is straightforward to check that the calculation for the $U(N)\times O(N)\times U(N)$ model of   \cite{Prakash:2017hwq} is essentially the same as for our model in \eqref{eq:general-int-M}. One finds the same anomalous dimension as in \eqref{eq:eta}, with $3\l_2^2$ replaced by the positive combination $\l^2=\l_1^2+\l_2^2-\l_1\l_2$, with $\l_1$ and $\l_2$ being now the two couplings of the tetrahedral interactions of the model in  \cite{Prakash:2017hwq}. Since the relation $\b_i=4\eta \l_i$ for $i=1,2$ still applies in the large-$N$ limit, both couplings are marginally irrelevant in two dimensions, and our conclusions for $d=2-\epsilon$ apply.}

\section{Intermediate field formalism for Dirac fermions}
\label{sec:Dirac-1color}

In the Dirac model, our study of the SD equation \eqref{eq:SD-D} suggests the existence of a subclass of theories in which the values of $\lambda_0^S$ and $\lambda_1^S$ conspire to give a vanishing non-perturbative mass \eqref{eq:dyn-mass-01}. In the present section, we will investigate further the viability of the massless condition \eqref{massless_cond}, by means of the intermediate field formalism (also known as Hubbard-Stratonovich transformation). As already apparent from the original GN literature \cite{Gross:1974jv}, this formalism is particularly useful to elucidate the stability properties of the tentative solutions to the SD equations. 

At first, it would seem that our tensorial version of GN necessitates the introduction of two types of intermediate fields: a) scalar fields associated to $I_0^X$ interactions, as in the standard GN model; and b) matrix fields associated to $I_1^X$ terms. However, it turns out that we can generate both $I_0^X$ and $I_1^X$ interactions with a single matrix intermediate field $M^X_{ij}$, provided we carefully adjust its free propagator. 

In order to simplify the discussion we consider a reduced model, keeping only the interactions $I_{0}^{S}$ and $I_{1}^{S,3}$ in  \eqref{eq:general-int-D}. This model is not color-invariant and it is not the most general one, but it captures the essential qualitative properties of the invariant one: we have seen in the previous section that the $P$ and $V$ interactions do not affect the SD equations; furthermore, as we will see below, the difference between the color symmetric and the not color-symmetric case simply amounts to some factors of~3. Note that such a truncation is stable under the renormalization group flow at leading order in $1/N$. In the language of \cite{Gurau:2011xp}, the $4$-point melonic diagrams generated by $I_{1}^{S,3}$ all have the same boundary graph: that of an $I_{1}^{S,3}$ interaction. This ensures that no effective $I_{1}^{S,\ell}$ interaction with $\ell \neq 3$ can be generated at leading order.

\bigskip

More explicitly, we consider in this section the reduced model:
\be
S =  S_{\rm free} + S'_{\rm int} \;,
\ee
with
\be
S'_{\rm int} = -\f{\l_0}{N^3} \int d^2 x\, \Tr[B]^2 - \f{\l_1}{N^2} \int d^2 x\, \Tr[B^2]\;,
\ee
where we have introduce the Hermitian matrix-valued bilinear
\be \label{eq:matrixB}
B_{ij} = (\psib_{a_1 a_2 i}  \psi_{a_1 a_2 j})\;.
\ee

We introduce the operator
\be
C_{ij;kl} = \delta_{ik} \delta_{jl}  - \frac{b}{N} \delta_{ij} \delta_{kl} 
\equiv ({\bf 1 - P})_{ij;kl}+ (1-b){\bf P}_{ij;kl} \;,
\ee
where 
\be \label{eq:P}
{\bf P}_{ij;kl}\equiv \frac{1}{N} \delta_{ij} \delta_{kl}
\ee
is the projector on the trace part of an $N\times N$ matrix, and thus ${\bf 1 - P}$ is the projector on its traceless part. 
The eigenvalues of $C$ are $(1 - b)$ with multiplicity 1 (eigenvector proportional to the identity matrix), and 1 with multiplicity $N^2 - 1$ (with eigenvactors the basis of traceless Hermitian matrices). Thus $C$ is always well defined, it is positive for $b < 1$ and it has a negative (zero) eigenvalue for $b > 1$ (respectively $b=1$). 
It is thus invertible for $b\neq 1$, with inverse
\be
C^{-1}= ({\bf 1 - P})+ \f{1}{1-b}{\bf P}\;.
\ee
Defining
\be \label{eq:b-def}
b= -\f{\l_0}{\l_1}\;,
\ee
we can rewrite the interaction action as
\be
S'_{\rm int} = - \f{\l_1}{N^2} \int d^2 x\; B^*_{ij}\, C_{ij;kl}\, B_{kl}\;.
\ee
By the standard Hubbard-Stratonovich trick we can then rewrite the exponent of the interaction action as an integral over a Hermitian matrix-valued field $M$ (the intermediate field); with obvious notation:
\be
e^{-S'_{\rm int}} = \int [\cD M] \, \exp\left\{ - \f12 \int d^2 x\, M^*\cdot C^{-1} \cdot M - \f{\sqrt{2\l_1}}{N}\int d^2 x\, \Tr[BM]\right\} \;.
\ee
Including also the fermions' free action, and making the notation completely transparent, the total Lagrangian density with the intermediate field reads
\be \label{eq:L-interm}
\tilde{\cL} [\psib , \psi , M] =  \psib_{abc} \slashed{\p} \psi_{abc} +  \frac{1}{2}  \left( \Tr[ M^2] + \frac{b}{(1 - b ) N} \Tr[M]^2 \right) + \f{\sqrt{2 \lambda_1}}{N} (\psib_{abi}  \psi_{abj}) M_{ij}\,.
\ee 
This action is still invariant under discrete chiral transformations, which now act as \eqref{eq:dct}, together with $M\to - M$.
As noticed above, for $b = 1$, $C$ becomes degenerate, which manifests itself by the divergence of the coefficient in front of $\Tr[M]^2$ in the Lagrangian.\footnote{The kinetic term becomes singular in this limit, but not the Gaussian measure itself, so that the path-integral remains well-defined.} The condition $b = 1$ is equivalent to the massless condition derived in the previous section, and is therefore particularly relevant. It amounts to imposing the traceless conditions $\Tr[ M ] = 0$ in the path-integral, as can be seen by applying the Hubbard-Stratonovich trick once more, this time to the double trace.
That is, by introducing a scalar $\phi$ and replacing
\be
\begin{split}
\exp &\left\{-\f12 \frac{b}{(1 - b ) N} \int d^2 x\, \Tr[M]^2 \right\} \\
& = \int [\cD \phi] \exp\left\{ - \frac{(1 - b ) N}{2}\int d^2 x\, \phi^2 + \sqrt{-b}\, \int d^2 x\, \phi\, \Tr[M] \right\}\,.
\end{split}
\ee
We see that for $b=0$ the field $\phi$ decouples from the rest of the action, while for $b=1$ the quadratic part vanishes and the integral over $\phi$ gives a Dirac delta for the trace of $M$.
We will use the representation with the field $\phi$ in appendix \ref{app:extrema}, while for the remainder of this section we will stick to the double-trace representation.

\subsection{Effective potential}

The $\psi$ fields can be integrated out from the partition function
\be
\begin{split}
Z &\propto \int  [\cD M] [\cD \psib \cD \psi] \exp\left( -\int d^2 x \tilde{\cL} [\psib , \psi , M]\right) \\
 \label{eq:Z}
& \propto \int [\cD M]  \exp\left( - \frac{1}{2} \int d^2 x  \left( \Tr[ M^2] + \frac{b}{(1 - b ) N} \Tr[M]^2 \right) + N^2 \widehat{\Tr} \ln \left[ R( M ) \right] \right) \,,
\end{split}
\ee
where
\be 
R( M) := 
 \slashed{\p}  + N^{-1} \sqrt{2 \lambda_1}   M \,,
\ee
and $\widehat{\Tr}$ includes a functional trace on top of the matrix trace.

To determine the vacuum configurations of the intermediate field, and hence check whether there is spontaneous symmetry breaking of the discrete chiral symmetry\footnote{See section \ref{sec:restoration} for more more details on symmetry breaking and symmetry restoration of the continuous $U(N)$ symmetry, in the light of the Coleman-Mermin-Wagner theorem.}, we want to compute the effective potential. By definition, the latter is the effective action (i.e.\ the generator of one-particle irreducible graphs) evaluated at constant field. In principle, in order to obtain the effective action we should introduce sources $J$ coupled to $M$, perform the functional integral and lastly do a Legendre transform of $W[J]\equiv \ln Z[J]$ with respect to the sources. 
However, by rescaling $M\to N M$ (and also $J\to N J$) we obtain
\be
\begin{split}
e^{W[J]}  &\propto  \int [\cD M]  \exp\left( - N^2 \int d^2 x  \left( \f12\Tr[ M^2] + \frac{b}{2(1 - b ) N} \Tr[M]^2 -\Tr[JM]\right) \right. \\
&\quad \left.+ N^2 \widehat{\Tr} \ln \left[  \slashed{\p}  +  \sqrt{2 \lambda_1}   M \right] \right) \\
& \equiv \int [\cD M]  \exp\left( -N^2 \int d^2 x\, \left(S[M]-\Tr[JM]\right)\right)\;.
\end{split}
\ee
and we see that the action is of order $N^3$ (each trace being a sum over $N$ elements).
As a consequence, in the large-$N$ limit we can bypass the integral over $M$ and evaluate $W[J]$ by saddle-point method, obtaining that it is simply given by the Legendre transform of $S[M]$.\footnote{One can show that the measure contribution leading to the standard Vandermonde determinant (and to eigenvalue repulsion) is subleading in $1/N$; we therefore ignore it.} 
And since the Legendre transform is an involution, we find that the effective action in the large-$N$ limit is simply given by $S[M]$ itself.
Therefore, the effective potential we are looking for is obtained as $V(M) = S[M(x)=M]/\int d^2 x$.

From now on we focus on constant fields $M$. The effective potential in Fourier space is then:
\be
V( M ) =  \frac{1}{2}  \left( \Tr[ (M )^2] + \frac{b}{(1 - b ) N} \Tr[M]^2 \right) -  \Tr \ln \left[ 1 -\im  \frac{\slashed{p}}{p^2}  \sqrt{2 \lambda_1}   M \right] \;.
\ee
When expanding the logarithm in power series, the fact that the trace of an odd number of gamma matrices vanishes implies that only positive powers survive, so that we get:
\be
\begin{split}
V(M) = & \frac{1}{2}   \left( \Tr[ M^2] + \frac{b}{(1 - b ) N} \Tr[M]^2 \right) \\
 &+  \sum_{n=1}^{+\infty} \frac{(-2 \lambda_1)^n}{n} \Tr[ M^{2n}] \int \frac{d^2 p}{(2 \pi)^2}  \left( \frac{1}{p^2} \right)^{n}  \,.
\end{split}
\ee
Including a UV regulator $\Lambda$ as well as an IR regulator at intermediate steps since, even though their sum is not, the amplitudes are also IR divergent, we obtain:
\begin{align}\nn
V( M) &= \frac{1}{2}   \left( \Tr[ M^2] + \frac{b}{(1 - b ) N} \Tr[M]^2 \right)  -  \int_{p^2 \leq \Lambda^2} \frac{d^2 p}{(2 \pi)^2} \Tr \ln \left( 1 + \frac{2 \lambda_1 }{p^2}   M^{2} \right) \\ \nn
&= \frac{1}{2}   \left( \Tr[ M^2] + \frac{b}{(1 - b ) N} \Tr[M]^2 \right) \\ \nn
& \quad - \frac{1 }{4 \pi } \left( 2 \lambda_1\Tr\left[   M^{2} \ln \left(1+ \frac{ \Lambda^2}{ 2 \lambda_1  M^{2}}\right) \right] +
\L^2 \Tr\ln \left( 1 + 2 \lambda_1( \frac{   M}{\L} )^{2} \right)  \right)   \\ \label{eq:potential}
&= \frac{1}{2}   \left( \Tr[ M^2] + \frac{b}{(1 - b ) N} \Tr[M]^2 \right)  + \frac{\lambda_1 }{2 \pi} \left( \Tr\left[  M^{2} \ln \frac{ 2 \lambda_1  M^{2}}{ \Lambda^2} \right] - \Tr[ M^{2} ]   \right)   \,.
\end{align}

Since the effective potential has a global $U(N)$ invariance, it is convenient to express it in terms of the eigenvalues $\mu_1, \ldots, \mu_N$ of $M$. We obtain:
\be 
V(M) = \sum_{i=1}^N v(\mu_i) + \frac{b}{2(1-b ) N} \left( \sum_{i=1}^{N} \mu_i \right)^2\,,
\ee
where
\be
v(\mu) := \frac{1}{2} \mu^2 + \frac{\lambda_1^S \mu^2}{2 \pi} \left( \ln\left(\frac{2 \lambda_1 \mu^2}{ \Lambda^2}\right) - 1 \right)
\ee
is nothing but the effective potential of the standard GN model \cite{Gross:1974jv}, up to a different normalization of the coupling constant.

\subsection{Beta functions}

The effective potential should satisfy as usual the renormalization group equation
\be
\left[  \L \f{\p}{\p\L} +\b_{\l_1} \f{\p}{\p\l_1} + \b_b \f{\p}{\p b} - \g \m_i \f{\p}{\p\m_i} \right] V = 0 \;,
\ee
from which we obtain
\be
\b_\l = 2\l\g = -\f{2}{\pi} \l_1^2 \;,
\ee
\be
\b_b = 2 \g b (1-b) \;.
\ee
Using \eqref{eq:b-def} we obtain also
\be
\b_{\l_0} = - \frac{2\lambda_0 ( \lambda_0 + 2 \lambda_1 )}{\pi} \;.
\ee
The beta functions so obtained for $\l_0$ and $\l_1$ are consistent with the ones obtained earlier for  $\l_0^S$ and $\l_1^S$ in \eqref{eq:beta0} and \eqref{eq:beta1}, up to extra factors of 3 due to the color permutations.

\subsection{Stationary points and their stability}
\label{sec:stability-Dirac}

Solutions to the equation of motion for the effective potential $V(M)$ (i.e.\ its stationary points) correspond to the vacuum expectation value of the field $M$ for different choices of vacua.
From \eqref{eq:potential}, the equations of motion read
\be \label{eq:Vprime}
0 = \f{\p V}{\p M_{ij}} = M_{ji} + \f{b}{(1-b)N} \d_{ij} \Tr[M] + \f{\l_1}{2\pi} \left\{M,\ln \f{2\l_1 M^2}{\L^2} \right\}_{ji} \;.
\ee
We can diagonalize them and rewrite
\be \label{eq:mu-eom}
\f{\p V}{\p M_{ij}}=0  \to \d_{ij} \left( \m_i +  \f{b}{(1-b)N} \sum_k \m_k + \f{\l_1}{\pi} \m_i \ln\f{2\l_1\m_i^2}{\L^2}  \right)=0\;,
\ee
where the variables $\m_i$, for $i=1,\ldots,N$, are the eigenvalues of $M$.

We seek a unitary invariant solution of the above equations of motion. The only $U(N)$-invariant matrix is a matrix proportional to the identity, hence we look for a solution such that $\m_i=\m$ for every $i$. Equation \eqref{eq:mu-eom} reduces to
\be
\m \Big(\f{1}{1-b} + \f{\l_1}{\pi} \ln \f{2\l_1\m^2}{\L^2} \Big) = 0 \;,
\ee
which admits the solutions $\m=0$ and
\be \label{eq:sym-mu}
\bar\m^2= \f{\L^2}{2\l_1} \exp\left(-\f{\pi}{\l_1(1-b)}\right)=  \f{\L^2}{2\l_1} \exp\left(-\f{\pi}{\l_0+\l_1}\right)\;.
\ee
The latter breaks the discrete chiral symmetry, while the former is of course invariant.

In order to investigate the stability of the solutions we need to look for the eigenvalues of the Hessian matrix $\f{\p^2V}{\p M_{ij} \p M_{kl}}$. The Hessian is difficult if not impossible to write at a general field configuration, since in general $M$ does not commute with its infinitesimal variation $\extd M$.\footnote{The problem lies in the expansion of $\ln(M+\extd M)$: we can rewrite it as $\ln(M(1+M^{-1}\extd M))$, but since $[M,(1+M^{-1}\extd M)]\neq 0$, and $M$ is not infinitesimal, we have an infinite series of terms of order $\extd M$, coming from the Baker-Campbell-Hausdorff formula.} We didn't have this problem when writing $\f{\p V}{\p M_{ij}}$ because of the trace in the potential, but once we take one derivative the trace is gone, and for the second derivative we encounter this problem.
However, the symmetric solution is proportional to the identity, so that it commutes with any matrix, and we can easily write the Hessian evaluated at the symmetric solution. It reads
\be
\f{\p^2V}{\p M_{ij} \p M_{kl}}_{\Big| M_{ab}=\m \d_{ab}} =
   \a(\m) ({\bf 1-P})_{ij;kl} + \b(\m)  {\bf P}_{ij;kl} \;,
\ee
where
\be
\a(\m) =  \left(1+ \f{2\l_1}{\pi} +\f{\l_1}{\pi}\ln\f{2\l_1\m^2}{\L^2}\right) \;,
\ee
\be
\b(\m) = \left(\f{1}{1-b}+ \f{2\l_1}{\pi} +\f{\l_1}{\pi}\ln\f{2\l_1\m^2}{\L^2}\right)\;,
\ee
and the operator ${\bf P}$ was defined in \eqref{eq:P}.
Therefore, the Hessian has eigenvalues $\a(\m)$ with multiplicity $N^2-1$, and $\b(\m)$ with multiplicity 1.
We can associate the latter with the variation of $\m$, since its eigenvector is proportional to the identity, while $\a(\m)$ is associated to infinitesimal $SU(N)$ transformations, its eigenvectors being given by the basis of traceless Hermitian matrices.

When $\mu \to 0$ both $\a(\m)$ and $\b(\m)$ go to $- \infty$, hence we conclude that the solution $\m=0$ is always unstable, just as in the GN model.

For the non-zero solution $\bar\m$ in equation \eqref{eq:sym-mu} we find instead
\be \label{eq:alpha-bar}
\a(\bar\m) =   \left( \f{2\l_1}{\pi} - \f{b}{1-b}\right) \;, \;\;\; \b(\bar\m)=\f{2\l_1}{\pi}\;.
\ee
In this case, all the eigenvalues are positive for $b < b_{c} \equiv \frac{2 \lambda_1}{\pi + 2 \lambda_1}$, hence the solution is stable in this range. Therefore, we have spontaneous breaking of the discrete chiral symmetry, with the generation of a mass $m$ for the fermions. The latter can be read off from \eqref{eq:L-interm} once we replace $M_{ij}\to N \bar{\m} \d_{ij}$:
\be \label{eq:dyn-mass-1col}
m = \sqrt{2\l_1} \bar\m = \pm \L \exp\left(-\f{\pi}{2(\l_0+\l_1)}\right) \;,
\ee
in agreement with \eqref{eq:dyn-mass-01}, again up to a factor 3 due to the color permutations.

However, since the $N^2-1$ degenerate eigenvalues $\a(\bar\m)$ become negative for $b > b_{c}$, we conclude that this vacuum becomes unstable before reaching the interesting value $b=1$, corresponding to $\lambda_0 + \lambda_1 =0$.
We might be tempted to interpret the appearance of zero-modes at $b=b_c$ as a signal of a second-order phase transition, but there are two problems with such an interpretation.
First, as we will see, at $b=0$ many disconnected global minima are present, and the symmetric solution with $\m_i=\bar\m$ ceases to be the global minimum at $b>0$. Therefore, we have a first-order phase transition at $b=0$. Second, and most important, any other minima break the $U(N)$ invariance of the model, but spontaneous symmetry breaking of a continuous symmetry is forbidden in two dimensions in virtue of the Coleman-Mermin-Wagner theorem. Therefore, we need further investigation in order to understand if the model at $b>0$ is consistent, and in case it is, whether after restoration of the $U(N)$ invariance it corresponds to a different phase (e.g. with no breaking of the discrete chiral symmetry) or not. We will come back to such question in section \ref{sec:restoration}.

The existence of other solutions is clear at $b=0$: the equations of motion \eqref{eq:mu-eom} decouple from each other, and reduce to 
\be \label{eq:b0eom}
 \m_i \left( 1 + \f{\l_1}{\pi} \ln\f{2\l_1\m_i^2}{\L^2}  \right)=0\;.
\ee
Therefore, besides the usual zero solution, we have
\be \label{eq:broken-mu}
\m_i =\pm \tilde\mu \equiv \pm \f{\L}{\sqrt{2\l_1}} \exp\left(-\f{\pi}{2\l_1}\right)\;,
\ee
where the plus or minus sign can be chosen independently for each $i=1,\ldots,N$. And since at $b=0$ the potential depends only on $\m_i^2$, for each choice of signs we obtain the same value of the potential. Hence we have $N+1$ minima of the potential, one for each choice of signs (up to permutations), two of which are the $b=0$ limit of \eqref{eq:sym-mu}, while all the others break the $U(N)$ invariance down to $U(N_+)\times U(N-N_+)$ (where $N_+$ is the number of plus signs in the given solution).

As we show in appendix \ref{app:extrema}, for $b\neq 0$ all the $N+1$ solutions above generalize to as many stationary points. However, the degeneracy gets lifted, and while for $b<0$ the symmetric solution \eqref{eq:sym-mu} is the global minimum, for $b>0$ it actually becomes the stationary point with the largest value of the potential. The global minimum at $b>0$ turns out to be the solution which maximizes the symmetry breaking, i.e. the one with an equal number of plus and minus signs (we assume from here on an even $N$).\footnote{When $N$ is odd, we can pick up $\frac{N-1}{2}$ positive (resp. negative) eigenvalues and $\frac{N+1}{2}$ negative (resp. positive) eigenvalues. The difference with respect to the $N$ even case is negligible in the large-$N$ limit.}
Such a solution exists at any $b$, and it is easily obtained by demanding the matrix $M$ to be traceless.
Then equations \eqref{eq:mu-eom} reduce again to \eqref{eq:b0eom}, with solution \eqref{eq:broken-mu}, plus the traceless constraint enforcing the number of plus and minus signs to be the same.

Evaluating the potential at the symmetric solution (equation \eqref{eq:sym-mu}) we get
\be \label{eq:v1}
v_1\equiv V(M_{ij}=\bar\mu\d_{ij}) = -\f{N\L^2}{4\pi} \exp\left(-\f{\pi}{\l_1(1-b)}\right)\;,
\ee
while at the maximally broken solution (equation \eqref{eq:broken-mu} with $N_+=N/2$) we get
\be \label{eq:v2}
v_2\equiv V(M_{ij}={\rm sgn}((N+1)/2-i)\tilde\mu\d_{ij}) = -\f{N\L^2}{4\pi} \exp\left(-\f{\pi}{\l_1}\right)\;.
\ee
At $b=0$ we have $v_1=v_2$ as discussed above, while $v_1<v_2$ for $b<0$ and $v_1>v_2$ for $b>0$, confirming that in the latter case the symmetric solution is no longer the global minimum.

To conclude the stability analysis, we can explicitly check that the maximally $U(N)$-breaking solution leads to the presence of $2N_+(N-N_+)=N^2/2$ (would-be) Goldstone modes. 
Since such solution has $\m_i^2=\tilde\m^2$ independent from $i$, we know that at this configuration $M^2$ (but not $M$ itself) commutes with $\extd M$, so we can expand the $\ln M^2$ that appears in the first derivative\footnote{That is, we use $\ln(\tilde{M}+dM)^2\simeq\ln(\tilde{M}^2+\tilde{M} dM+ dM \tilde{M}) = \ln \tilde{M}^2 + \ln (1+\tilde{M}^{-2}(\tilde{M} dM+ dM \tilde{M}))$, which holds thanks to the fact that $\tilde{M}^2$ is proportional to the identity.} and obtain:
\be
\f{\p^2V}{\p M_{ij} \p M_{kl}}_{\Big| M_{ab}={\rm sgn}((N+1)/2-a)\tilde\mu \d_{ab}} =
   \tilde\a_{kl}(\tilde\m) ({\bf 1-P})_{ij;kl} +\tilde\b(\tilde\m)  {\bf P}_{ij;kl} \;,
\ee
where
\be \label{eq:alphatilde}
\tilde\a_{kl}(\tilde\m) =   \f{\l_1}{\pi} (1+{\rm sgn}(\m_k){\rm sgn}(\m_l)) \;, \;\;\; \tilde\b(\tilde\m)=\f{2\l_1}{\pi}+\f{b}{1-b}\;.
\ee
We see that $\tilde\a_{kl}(\tilde\m)=0$ for 
$k\leq N/2$ and $l > N/2$, as well as for $l\leq N/2$ and $k > N/2$;
 hence we have $N^2/2$  massless modes.
These look like the Goldstone modes associated to the spontaneous symmetry breaking of $U(N)$ down to $U(N/2)\times U(N/2)$. However, spontaneous symmetry breaking of a continuous symmetry is forbidden in two dimensions \cite{Coleman:1973ci, Mermin:1966fe}, therefore we will refer to them as ``would-be Goldstone modes'', and we will discuss their role in section~\ref{sec:restoration}.

\subsection{2-point function}

The effective action for the intermediate field generates by definition the one-particle irreducible (1PI) $n$-point functions of the intermediate field. And since the latter is conjugated to a fermionic bilinear, one can say that the effective action for the intermediate field generates the 1PI $n$-point functions for that particular class of composite fields (see \cite{Gross:1974jv} for the precise relation in the GN model case).
Here we compute the $2$-point function first in the symmetric vacuum, and then in the maximally-broken one. For the former, we show that the instability at $b>b_c$ manifests itself in the presence of tachyonic poles. In the second case, we show that at small momentum $p$ the behavior of the 2-point function is the standard $1/p^2$, corresponding to a logarithm in position space, the consequence of which we will discuss in the following subsection.

The computation of the 2-point function is most easily performed using the action with the fermion fields not yet integrated out, i.e. from the Lagrangian \eqref{eq:L-interm}, expanded either around the stationary point \eqref{eq:sym-mu} or  \eqref{eq:broken-mu} with $N_+=N/2$.

\begin{figure}[ht]
 \centering
             \includegraphics[scale = 0.75]{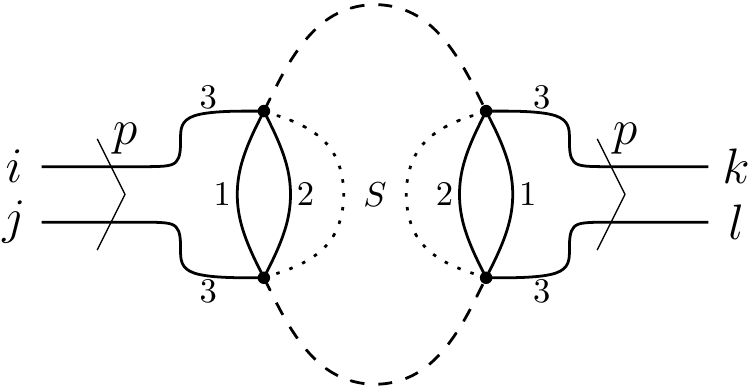}
  \caption{One-loop self energy $\Pi_{ij;kl}(p)$ of the intermediate matrix field.}
\label{fig:matrix_self}
\end{figure}

We define the intermediate field connected $2$-point function (or propagator) as
\be
D_{ij;kl} = \langle M_{ij} M^*_{kl} \rangle_{\rm conn} \;,
\ee
where the average is taken on the vacuum state. Its inverse, i.e.\ the second variation of the effective action at the field configuration corresponding to the vacuum state, satisfies the usual SD equation
\be
D^{-1}_{ij;kl} = C^{-1}_{ij;kl} - \Pi_{ij;kl} \;,
\ee
where $\Pi$ is the self-energy. The latter is given in the large-$N$ limit by a single diagram, with a single fermionic loop, shown in figure \ref{fig:matrix_self}.

In the symmetric vacuum, the self-energy in momentum space is given by
\be
\Pi_{ij;kl}(p) = - 2\l_1 \int \f{d^2 q}{(2\pi)^2} \Tr\left[ \left(\f{-\im\slashed{q}+m}{q^2+m^2}\right) \left(\f{-\im(\slashed{q}-\slashed{p})+m}{(q-p)^2+m^2} \right) \right] \d_{ik} \d_{jl}
 \equiv - 2\l_1 I(p) \d_{ik} \d_{jl}\;,
\ee
where we have used the massive fermionc propagator, with mass \eqref{eq:dyn-mass-1col}.
The integral $I(p)$ is UV divergent and thus requires a UV cutoff $\L$. After a standard computation we obtain
\be
I_\L(p) = -\f{1}{2\pi} \ln\left(1+\f{\L^2}{m^2}\right) + \f{1}{2\pi} B(p,m)\;,
\ee
where we defined as in \cite{Gross:1974jv}
\be
B(p,m) \equiv \f{\sqrt{p^2+4m^2}}{p}\; \ln\f{p+\sqrt{p^2+4m^2}}{-p+\sqrt{p^2+4m^2}} \;.
\ee
Imposing the renormalization condition
\be
D^{-1}_{ij;kl}(p=0) = \f{\p^2V}{\p M_{ij} \p M_{kl}}_{\Big| M_{ab}=\bar\m \d_{ab}}\;,
\ee
we obtain at last
\be
D_{ij;kl} =  f(p)^{-1} ({\bf 1-P})_{ij;kl} 
+ \frac{\pi}{\lambda_1 B(p,m)}  {\bf P}_{ij;kl} \;,
\ee
where 
\be
f(p) = \a(\bar\m)+\f{\l_1}{\pi} (B(p,m)-2) = - \f{b}{1-b} +\f{\l_1}{\pi} B(p,m)\;.
\ee
Since $B(p,m)$ is a monotonically increasing function of $p$, such that $B(0,m)=2$ and at large momentum $B(p,m) = \ln\f{p^2}{m^2} +O(\f{1}{p^2})$, we see that as soon as 
$\frac{2 \lambda_1}{\pi} < \frac{b}{1-b}$ (i.e. for $b>b_c$) the function $f(p)^{-1}$ has a pole at $p>0$. Since we are in Euclidean signature, the latter is a tachyon.

Let us now consider the connected two-point function on the maximally broken vacuum. The calculation goes as before, except for the renormalization condition at zero momentum, which this time should reproduce the Hessian around the symmetry-breaking vacuum, and for the fact that we need to keep track of the different signs in the mass terms. 

For the self-energy we have
\be \label{eq:selfE-broken}
\tilde\Pi_{ij;kl}(p) =  - 2\l_1 I_\L^{kl}(p) \d_{ik} \d_{jl}\;,
\ee
with
\be 
I_\L^{kl}(p) = 2 \int \f{d^2 q}{(2\pi)^2} \f{\tilde{m}^2 \mathrm{sgn}(\mu_k) \mathrm{sgn}(\mu_l) - q\cdot (q-p)}{(q^2 + \tilde{m}^2) [(q-p)^2 + \tilde{m}^2]}\,. 
\ee

We obtain
\be
D_{ij;kl}^{-1} =  \tilde f_{kl}(p) ({\bf 1-P})_{ij;kl} + \left(\tilde f_{kk}(p)+\f{b}{1-b}\right)  {\bf P}_{ij;kl} \;,
\ee
where
\be
\tilde f_{kl}(p) = \tilde\a_{kl}(\tilde\m)+\f{\l_1}{\pi} (\tilde B_{kl}(p,\tilde{m})-(1+\mathrm{sgn}(\mu_k) \mathrm{sgn}(\mu_l)))\;,
\ee
with $\tilde\a_{kl}(\tilde\m)$ given in \eqref{eq:alphatilde}, and with
\be
\tilde B_{kl}(p,m) \equiv \f{p^2+2m^2(1+\mathrm{sgn}(\mu_k) \mathrm{sgn}(\mu_l))}{p\sqrt{p^2+4m^2}} \ln\f{p+\sqrt{p^2+4m^2}}{-p+\sqrt{p^2+4m^2}} \;.
\ee
Since $\tilde f_{kl}(p) =\f{\l_1}{\pi} \tilde B_{kl}(p,\tilde{m})$ for $k\leq N/2$ and $l > N/2$, as well as for $l\leq N/2$ and $k > N/2$, 
and since at small momentum $\tilde B_{kl}(p,m)\simeq p^2/(2m^2)$, for the modes in such siubspace we obtain the standard behavior of a massless boson, which in coordinate space corresponds to a logarithmic divergence at large $x$.

\subsection{Effective non-linear sigma model and Coleman-Mermin-Wagner theorem}
\label{sec:restoration}

The intermediate field analysis of the Dirac model shows that the symmetric solution anticipated from the SD equation in section \ref{sec:sde} is unstable in the limit $b \to 1$ (i.e. $\lambda_0 + \lambda_1 = 0$). The true vacuum configuration of the intermediate field generates a non-zero mass and thereby breaks the discrete chiral symmetry of the bare action. More surprisingly, it also seems to break the continuous $U(N)$ symmetry of the model, which would be in obvious tension with the Coleman-Mermin-Wagner theorem \cite{Mermin:1966fe,Coleman:1973ci}.

This is reminiscent of a similar conundrum arising in the chiral version of the GN model \cite{Gross:1974jv}: in this theory, the discrete chiral symmetry is enhanced to a global $U(1)$ symmetry, which looks naively broken by the spontaneously generated mass. However, it was shown by Witten that this impression follows from the incorrect assumption that the chiral angle admits a well-defined expectation value in the broken vacuum \cite{Witten:1978qu} (see also \cite{Anninos:2010sq} for a recent discussion of this problem in the context of the AdS/CFT correspondence). Similarly, we will show that the $U(N)$ angles parametrizing the broken solutions of our model have a non-trivial dynamics governed by an effective non-linear sigma model, which suggests that a similar mechanism of symmetry restoration as in the chiral GN model is at play. In view of the involved dynamics of the massless modes we will uncover, we will not be able to provide a completely explicit check of symmetry restoration in our theory, but analogies with the chiral GN model will allow us to conclude with confidence that it does indeed occur.   

Before moving to the specifics of our model, let us describe the main mechanism behind symmetry restoration in two dimensions. Consider a theory with a continuous global symmetry, and a set of covariant operators $\cO_A$ whose expectation values are zero by symmetry, but such that some linear combination of their 2-point functions $C_{AB} \la \cO_A(x) \cO_B(0) \ra$ can be non-zero (because $C_{AB} \cO_A(x) \cO_B(0)$ is an invariant under global transformations). By the cluster decomposition theorem, if the symmetry is not broken by the vacuum (i.e. if the one point functions remain zero) then the 2-point function must satisfy
\be
\lim_{|x|\to\infty} C_{AB} \la \cO_A(x) \cO_B(0) \ra = C_{AB} \la \cO_A(x)\ra \la \cO_B(0) \ra = 0\;.
\ee
On the contrary, if the limit is non-zero we have spontaneous symmetry breaking (with $ \la \cO_A\ra = \r_A\neq 0$). In such case, by Goldstone's theorem there must be massless modes corresponding to the broken symmetries. In the deep IR we expect only the latter to be relevant, so it makes sense to freeze all the other (massive) modes and only look at the massless fluctuations around the vacuum. In dimension $d>2$ these lead to 
\be
\lim_{|x|\to\infty} C_{AB} \la \cO_A(x) \cO_B(0) \ra \simeq C_{AB} \r_A \r_B + \f{\k}{|x|^{d-2}}\;,
\ee
consistently with the symmetry breaking picture (with the constant piece coming from the disconnected contribution).
However, in $d=2$ we obtain 
\be \label{eq:2pt-log}
\lim_{|x|\to\infty} C_{AB} \la \cO_A(x) \cO_B(0) \ra \simeq C_{AB} \r_A \r_B  + \k \ln|x|\;,
\ee
and since the logarithm diverges at large distances the expansion breaks down. What this means is that the massless modes fluctuate wildly at large distances and they cannot be treated perturbatively. Their nonperturbative treatment will generally lead to symmetry restoration (for example this happens if in \eqref{eq:2pt-log} one has actually expanded $|x|^{-\a} = e^{-\a \ln |x|}$ to first order in $\a$, which of course gives a wrong result at large $|x|$). We would like to explicitly check this scenario for our model at $b>0$.

We have seen in the previous subsection that indeed in the broken vacuum the would-be Goldstone modes lead to a logarithmic divergence for the (connected) 2-point functions at large $x$.
However, this comes from having chosen one (simple) representative in the space of degenerate vacua and having expanded the unitary matrices around the identity. We want to keep them nonperturbative (or in other words average over the vacua) and compute
\be\label{2point_nlsm}
\begin{split}
K(x) &\equiv \Tr \langle  M(x)  M(0) \rangle \\
&\simeq \Tr \langle U^\dagger(\theta(x)) \tilde M U(\theta(x)) U^\dagger(\theta(0)) \tilde M U(\theta(0)) \rangle \;,
\end{split}
\ee
where $\tilde M_{ab}={\rm sgn}((N+1)/2-a)\tilde\mu \d_{ab}$, and $\theta(x)$ stands for the ($x$-dependent) $N^2-N$ angles parametrizing $U(N)/U(1)^{\otimes N}$.
In the second line we have frozen the $N$ modes corresponding to shifts in the eigenvalues, which are massive and thus irrelevant in the deep IR.
Furthermore, since $\tilde M_{ab}$ is invariant under the subgroup $U(N/2)\times U(N/2)$ (the stabilizer), we only need to consider unitary matrices in the coset space $\mathrm{Gr}(N,N/2)\equiv U(N)/(U(N/2)\times U(N/2))$, also known as the complex Grassmannian space.

By symmetry arguments, the low energy effective action for the would-be Goldstone modes $V \in \mathrm{Gr}(N,N/2)$ must be given by a complex Grassmannian non-linear sigma model 
\be\label{eq:NLSM}
\cL_{\rm eff} \propto \Tr\left[ \p_\m V^\dagger \p_\m V \right]\;.
\ee
Let us investigate this point in more detail and determine the proportionality factor in equation \eqref{eq:NLSM}. 

As previously argued, in order to capture the IR massless modes, we may restrict the path-integral over $M$ to the domain:
\be 
F := \{ U \tilde{M} U^{\dagger} \vert U \in U(N) \} \simeq U(N)/(U(N/2) \times U(N/2)) 
\ee
Following the literature (see e.g. \cite{Eichenherr:1979ci, Brezin:1979am, Hikami:1978ya}), it is convenient to parametrize $F$ in terms of orthogonal projectors $P$ of fixed rank $\mathrm{rk}(P) =N/2$. This is simply given by the change of variables
\be
M(x) = \tilde{\mu} \left( 2 P(x) - 1 \right)
\ee
To explicitly check the nature of $P$, it suffices to note that
\be
P(x) = \frac{1}{2 \tilde{\mu}}\left( M(x) + \tilde{\mu} \right) = U(x) \tilde{P}  U^\dagger(x) \,, \qquad \tilde{P} := \begin{pmatrix}   \one & 0 \\ 0 & 0 \end{pmatrix}
\ee
When $U$ is varied over $U(N)$, $P$ spans the set of all orthogonal projectors of rank $N/2$. Hence the integration over the fields $M(x)$ can be replaced by an integration over operators $P(x)$ verifying: $P^2 = P = P^\dagger$ and $\Tr P = N/2$. Moreover, the Jacobian of this transformation is independent of $x$ and can therefore be ignored. 

Consider a fixed configuration $P(x) = U(x) \tilde{P}  U^\dagger(x)$ of the projector field. There clearly is some freedom in our choice of matrix field $U$: we can multiply it on the right by unitaries which commute with $\tilde{P}$, and on the left by unitaries which commute with $P$. By differentiation of the relation defining $P$, one finds that:
\be\label{flow_p}
\extd P = \left[ P , U dU^{\dagger}\right]
\ee

The matrix $P$ being Hermitian, the operator $\hat{P}: A \mapsto \left[ P , A \right]$ is itself Hermitian on the space of $N\times N$ matrices equipped with the inner product $\langle A , B\rangle = \Tr( A^\dagger B )$. It has eigenvalues $0$, $2$ and $- 2$, with multiplicities  $N^2 /2$, $N^2 / 4$ and $N^2 / 4$ respectively. Hence the Maurer-Cartan form $\extd \tilde{A} := U \extd U^\dagger$ decomposes uniquely into $\extd\tilde{A} = \extd\tilde{A}_{0} + \extd\tilde{A}_{+} + \extd\tilde{A}_{-}$, where $\extd \tilde{A}_{0}$ is the orthogonal projection on the null eigenspace of $\hat{M}$ and $\extd\tilde{A}_{\pm}$ are the projections on the eigenspaces with eigenvalues $\pm 2 \tilde{\mu}$. Since $\extd\tilde{A}_0$ does not contribute to the flow of $P$ \eqref{flow_p}, one can fix the gauge (up to a global transformation) by the condition:
\be
\extd\tilde{A}_{0} = 0 \quad \Leftrightarrow \quad \forall x, \; \p_\mu \tilde{A}_{0}(x) = 0
\ee 
More explicitly, the null eigenspace of $\hat{P}$ is immediately seen to be $E_{0}(P)=P\cM_N(\mathbb{C})P \oplus (1-P)\cM_N(\mathbb{C})(1-P)$, with $\cM_N(\mathbb{C})$ being the space of complex $N\times N$ matrices, and the gauge-fixing condition is therefore equivalent to:
\be
P U \extd U^{\dagger} P = 0 \qquad \mathrm{and} \qquad (1-P) U \extd U^{\dagger} (1-P) = 0
\ee
In turn, these two equations are equivalent to:
\be 
\tilde{P} U^{\dagger} \extd U \tilde{P} \qquad \mathrm{and} \qquad (1-\tilde{P}) U^{\dagger} \extd U (1-\tilde{P}) = 0
\ee
We therefore conclude that, in the chosen gauge, the Maurer-Cartan form $\extd A := U^\dagger \extd U$ takes the form
\be\label{udu}
\extd A =  \im \begin{pmatrix}   0 &  \extd B \\  \extd B^\dagger & 0 \end{pmatrix}
\ee
for some $B$. Another useful relation is obtained by conjugation of the flow equation \eqref{flow_p}:
\be\label{dP_rot}
U^{\dagger} \extd P U = \left[ \extd A , \tilde{P} \right] = \im \begin{pmatrix}   0 &  - \extd B \\  \extd B^\dagger & 0 \end{pmatrix}
\ee

We can now proceed with the computation of the effective kinetic term of $P$, starting from the Lagrangian:
\be
\cL_{\mathrm{IR}} = \psib \slashed{\p} \psi + \tilde{m} \psib_{ibc} \psi_{jbc} \left( 2 P_{ij} - 1 \right)
\ee
We express $P(x) = U(x) \tilde{P} U^\dagger (x)$ with $U \extd U^{\dagger} \in E_0 (P)$, and change variables in the path-integral over the fermions: $\psi(x) \to U(x) \psi(x)$ and $\psib(x) \to \psib(x)  U^{\dagger}(x)$. The path-integral measure is invariant under this local unitary transformation, but the kinetic term is not and generates an effective coupling between the Maurer-Cartan form and the the fermions:
\be
\cL_{\mathrm{IR}}' = \psib_{ibc} \left( \slashed{\p} + \mathrm{sgn}(\tilde{\mu}_i) \tilde{m} \right) \psi_{ibc} + \psib_{ibc} \gamma^\mu \psi_{jbc} \partial_\mu A_{ij}\,.
\ee  
Importantly, our construction ensures that $\partial_\mu A := U^{\dagger} \partial_\mu U$ is of the block-diagonal form \eqref{udu}.

Integrating out the fermions we generate as before an effective action, this time for the Maurer-Cartan form. Since $\p_\m A_{ij}$ has dimension one, and since no terms in $A_{ij}$ without derivatives can be generated (because of the original global symmetry of the action), we are interested just in the part of the action which is quadratic in $\p_\m A_{ij}$ (with no additional derivatives), higher order terms being irrelevant in the IR (and the linear part being zero). 
This is given by the same type of diagrams in figure \ref{fig:matrix_self} we encountered before for the self-energy of $M$ (in the maximally broken vacuum). That is, we have
\be 
\cL_{\rm eff} = -\f{1}{2} \partial_\m A_{ij} \partial_\n A_{lk} \hat\Pi^{\m\n}_{ij;kl}(0)\;,
\ee
with
\be
\hat\Pi^{\m\n}_{ij;kl}(p) = - N^2 \int \f{d^2 q}{(2\pi)^2} \Tr\left[ \g^\m \left(\f{-\im\slashed{q}+\tilde{m}\,\mathrm{sgn}(\mu_k)}{q^2+\tilde{m}^2}\right)\g^\n \left(\f{-\im(\slashed{q}-\slashed{p})+\tilde{m}\,\mathrm{sgn}(\mu_l)}{(q-p)^2+\tilde{m}^2} \right) \right] \d_{ik} \d_{jl}
 \;.
\ee
Using
\be
\Tr(\g^\m \g^\r \g^\n \g^\s) = 2 \d_{\m\r} \d_{\n\s} - 2 \eps_{\m\r} \eps_{\n\s} =  2 (\d_{\m\r} \d_{\n\s} +  \d_{\m\s} \d_{\n\r} -  \d_{\m\n} \d_{\r\s})\;,
\ee
and the fact that any trace of an odd number of gamma matrices is zero, we obtain
\be
\hat\Pi^{\m\n}_{ij;kl}(0) = -N^2 \f{1}{2\pi} \f{\L^2}{\tilde{m}^2+\L^2}
 \mathrm{sgn}(\mu_k) \mathrm{sgn}(\mu_l) \d^{\m\n} \d_{ik} \d_{jl}\;.
\ee
Since $\tilde{m}$ is $\L$-independent, we can take the limit $\L\to+\infty$ at fixed $\tilde{m}$, and obtain
\be
\hat\Pi^{\m\n}_{ij;kl}(0) =-N^2 \f{1}{2\pi} \mathrm{sgn}(\mu_k) \mathrm{sgn}(\mu_l)\d^{\m\n} \d_{ik} \d_{jl}\;.
\ee

Remembering that our gauge-fixing condition ensures a block-diagonal form of $\extd A$, see \eqref{udu},  we obtain:
\be
\cL_{\mathrm{eff}} = \f{-N^2}{4 \pi} \Tr\left[\p_\m A \p^\m A\right]\,. 
\ee
We can finally use \eqref{udu} and \eqref{dP_rot} to conclude that:
\be\label{nlsm_action}
\cL_{\mathrm{eff}} = \f{-N^2}{4 \pi} \Tr\left[\p_\m A \p^\m A\right] = \f{N^2}{2 \pi} \Tr\left[\p_\m B \p^\m B^\dagger\right] = \f{N^2}{2 \pi} \Tr\left[\p_\m P \p^\m P\right]
\ee
The key observation to be made is that the explicit dependence in our choice of unitary matrix $U$ drops out, as it should. Remark also that \eqref{nlsm_action} is mapped to \eqref{eq:NLSM} after identifying $V(x) = 2 P(x) - 1$.

In order to explicitly check the restoration of the $U(N)$ symmetry at the level of the two-point function \eqref{2point_nlsm}, we would need to determine the large $\vert  x \vert$ asymptotics of:
\be\label{eq:V-2pt}
K(x) = 
\tilde{\mu}^2 \Tr \langle (2P(x) - 1) (2P(0) - 1) \rangle = 4 \tilde{\mu}^2 \Tr \langle P(x) P(0) \rangle - \tilde{\mu}^2 N\,.
\ee
Notice the $N^2$ factor in front of the action \eqref{nlsm_action}. As discussed before, this marks a difference with usual matrix-valued models, which would have just a factor $N$, and for which the Vandermonde determinant would thus play an important role. In our case, the large-$N$ limit at leading order tells us simply that $\p_\m V = 0$, i.e. $V$ is constant. In such case, in evaluating \eqref{eq:V-2pt} at leading order we find $K(x) = N \tilde{\m}^2$, and therefore we would seem to have spontaneous symmetry breaking of the $U(N)$ invariance. In the case of the apparent symmetry breaking of continuous chiral symmetry in the GN model, Witten determined that the two-point function of interest goes like $|x|^{-1/N}$. This observation explains why one may find a non-zero constant in the limit $N \to \infty$, while at the same time obeying the Coleman-Mermin-Wagner theorem. In particular, the $U(1)$ chiral symmetry is preserved, and though there is an infrared massless mode in the theory (the so-called $\theta$ particle), it is not a Goldstone boson. By analogy, in the present model we might expect to find a power-law decay at finite $N$ and large $|x|$, $K(x)\sim |x|^{-c/N^\a}$ for some constant $c$ and a positive exponent $\a$, corresponding to quasi-long-range order. However, to the best of our knowledge there are no examples of this behavior for non-abelian symmetries, and on the basis of the generic features of non-linear sigma models on homogenous spaces, such as asymptotic freedom and dynamical mass generation, we expect to find more likely an exponential decay, $K(x)\sim e^{-m|x|/N^\a}$, corresponding to disorder.\footnote{One should not get confused by the fact that the expansion of the exponential for large $N$ does not lead to a $\ln|x|$, in apparent contradiction with our discussion above: the exponential decay is only valid at finite $N$ and large $|x|$. For both $|x|$ and $N$ finite, the massive propagator is a complicated function of $m|x|/N^\a$ (e.g. a Bessel function in the free scalar case) and the limits of large $|x|$ and large $N$ do not commute: the former leads to an exponential decay, while the latter leads to a logarithmic behavior.}
Checking this explicitly in the complex Grassmannian sigma model \eqref{nlsm_action} is however technically challenging, and certainly goes beyond the scope of the paper. We are not aware of any completely conclusive calculation in this respect, but we refer the reader to \cite{McKane:1979cm} for an early attempt based on a renormalization group analysis.

\section{Conclusion and outlook}\label{sec:conclu}

We have studied the large-$N$ limit of a new class of fermionic models in two dimensions with quartic interaction and a tensorial index structure. 

In many respects the Dirac models with $U(N)^3$ symmetry turn out to be still very similar to the usual Gross-Neveu model: they are asymptotically free and they have a non-perturbative mass gap. 
For such models  we have introduced an intermediate matrix field representation, by means of which we have discovered a first order phase transition to a phase with apparent breaking of the $U(N)^3$ symmetry, and with an effective action for the would-be Goldstone modes given by a complex Grassmannian non-linear sigma model. The $U(N)^3$ symmetry is non-perturbatively restored in two dimensions, as we discussed in section \ref{sec:restoration}, but it would be interesting to explore such new phase in higher dimensions.

The Majorana models with $O(N)^3$ symmetry represent a major departure from GN-type physics. They are in fact the models which most closely resemble the SYK model, satisfying for example similar large-$N$ Schwinger-Dyson equations; the latter imply in particular that their self-energy has a non-trivial momentum dependence, thus signaling a crucial departure from the standard GN model. However, they have a crucial difference also with respect to the SYK model: in two dimensions the coupling is marginal and the free-propagator term in the Schwinger-Dyson equation cannot be discarded neither in the IR nor in the UV.
It turns out that at leading order in $1/N$ the coupling of their tetrahedral interaction is marginally irrelevant, i.e. the free theory is UV unstable. As a consequence, the theory develops a weakly interacting IR fixed point in $d=2-\epsilon$ dimensions for small $\epsilon$, as recently  conjectured in \cite{Prakash:2017hwq}.
In precisely two dimensions we found no non-trivial conformal field theories within our space of theories.

One of the most appealing features of the Gross-Neveu models, together with their large-$N$ properties, is that they are \textit{integrable}, see e.g.~\cite{Abdalla:1991vua,Bombardelli:2016rwb}. This means that, as classical field theories, they possess an infinite number of conserved charges which constrain their dynamics~\cite{deVega:1984wk,Hauer:1997ig}. Remarkably, integrability carries over to the quantum level, taking the form of Yangian symmetry, see for example \cite{Loebbert:2016cdm}; this allows to fix the quantum scattering matrix of the theory exactly, even at finite~$N$. From this point of view, tensorial Gross-Neveu model are an ideal playground for exploring integrability for $1+1$-dimensional tensor models. In fact, for the simplest interaction which we considered, $I_0^S$, our model simply gives several copies of the usual GN model, and integrability should follow automatically, albeit somewhat trivially. Exploring more general interactions, however, might yield integrable deformations of the underlying Yangian algebra. It would be interesting to investigate this possibility, and in particular to elucidate whether there is any relation between integrability and color symmetry.

Another possible avenue for further applications of tensor models are large $N$ gauge theories. In $d=1$ models, gauging the $G^{r}$ symmetry is straightforward as it essentially amounts to restricting the set of observables to the singlet sector \cite{KlebanovTarnopolsky}. In contrast, the gauge potential acquires a non-trivial dynamics in higher dimensions, which might allow to construct interesting large-$N$ tensorial gauge theories.

\section*{Acknowledgements}

\noindent D.~B.~ would like to thank Vincent Rivasseau and A.~S.~would like to thank Ben Hoare for helpful discussions.

\noindent This research was supported in part by Perimeter Institute for Theoretical Physics. Research at Perimeter Institute is supported by the Government of Canada through the Department of Innovation, Science and Economic Development Canada and by the Province of Ontario through the Ministry of Research, Innovation and Science.

\noindent A.~Sfondrini acknowledges support by the ETH ``Career Seed Grant'' n.~0-20313-17, as well as partial support from the NCCR SwissMAP, funded by the Swiss National Science Foundation.

\newpage
\appendix
\section{Euclidean fermions: conventions and useful formulas}
\label{app:conventions}

We work in Euclidean signature, i.e. $g_{\m\n}=\d_{\m\n}\equiv {\rm diag}(1,1)$.
The Levi-Civita tensor $\eps_{\m\n}$ is chosen such that $\eps_{12}=-\eps_{21}=1$.

In two spacetime dimensions the Clifford algebra $\{\g_\m,\g_\n\}=2\d_{\m\n}$ admits two-dimensional representations.
In fact, it suffices to notice that the Pauli matrices $\s_x$, $\s_y$, and  $\s_z$ satisfy the three-dimensional Euclidean Clifford algebra, hence one can select two of them as $\g_\m$, and the third one as $\g_5$ (the subscript ``5'' really makes sense only in four dimensions, but we follow the very common convention of using the same notation also in two dimensions).
We will mostly work in the following Majorana representation:
\be
\g_1 = \s_x = \begin{pmatrix} 0 & 1 \\ 1 & 0 \end{pmatrix} \;, \;\;\; 
 \g_2 = \s_z = \begin{pmatrix} 1 &  0 \\ 0 & -1 \end{pmatrix}  \;,
\ee
\be
\g_5 = -\s_y = \begin{pmatrix} 0 & \im  \\  -\im & 0 \end{pmatrix} = -\im \g_1 \g_2  \; .
\ee

In the case of Dirac fermions the following Weyl representation can also be useful:
\be
\g_1 = \s_y = \begin{pmatrix} 0 & - \im \\ \im & 0 \end{pmatrix} \;, \;\;\; 
 \g_2 = -\s_x = \begin{pmatrix} 0 &  -1 \\ -1 & 0 \end{pmatrix}  \;,
\ee
\be
\g_5 = \s_z = \begin{pmatrix} 1 & 0  \\  0 & -1 \end{pmatrix} = -\im \g_1 \g_2 = -\g^0 \g^1 \; .
\ee
For both representations we have the identities: 
\be
[\g_\m,\g_\n] = 2 \im \eps_{\m\n} \g_5\;,
\ee
\be
[\g_\m , \g_5] = -2 \im \eps_{\m\n} \g_\n\;,
\ee
\be
\{\g_\m,\g_5\}=0 \;,
\ee
\be
\g_\m \g_\n \g_\r = \d_{\m\n} \g_\r - \eps_{\m\n} \eps_{\r\s} \g_\s\;.
\ee
Furthermore, both $\gamma_\m$ and $\gamma_5$ are Hermitian matrices and they square to the identity matrix.

We denote the spinorial indices by capital letters from the middle of the latin alphabet, e.g. $I,J,K,L=1,2$.
We have the following useful relation:
\be \label{eq:eps-delta}
\eps_{IJ}\, \eps_{KL} = \d_{IK} \d_{JL} - \d_{IL} \d_{JK} \; ,
\ee
and the completeness relation
\be \label{eq:compl}
\d_{IL} \d_{KJ}  = \f12\big( \d_{IJ} \d_{KL}+ (\g_\m)_{IJ} (\g_\m)_{KL} + (\g_5)_{IJ} (\g_5)_{KL} \big)\;.
\ee
By contracting the latter with either two $\g_5$ or two $\g_\m$ we find also
\begin{align} \label{eq:compl-2}
(\g_5)_{IL} (\g_5)_{KJ}  &= \f12\big( \d_{IJ} \d_{KL} -  (\g_\m)_{IJ} (\g_\m)_{KL} +  (\g_5)_{IJ} (\g_5)_{KL}\big)\,, \\
 \label{eq:compl-3}
(\g_\m)_{IL} (\g_\m)_{KJ}  &= \d_{IJ} \d_{KL} - (\g_5)_{IJ} (\g_5)_{KL}\,.
\end{align}

For what concerns us, Majorana (Dirac) fermions $\psi_{\bf a}^I$ are real\footnote{More precisely, Majorana fermions are only real in the Majorana representation, otherwise they satisfy a generalized reality condition.} (complex) Grassmann fields, with a spinorial index $I$ labelling the two spinorial components.
The boldface index ${\bf a}$ denotes a collective flavor index associated to a representation of the symmetry group $G$ of the model (for example ${\bf a}\equiv i=1,\ldots,N$ in the $O(N)$-invariant GN model, or ${\bf a}\equiv abc$, with $a,b,c=1,\ldots,N$ for the $O(N)^3$-invariant tensorial version).
We will usually omit one or the other type of indices, unless they are necessary.

The definition of  $\psib$ is a bit tricky in Euclidean signature, and it differs significantly in the literature. 
One option, followed by many, is to double the number of spinor fields in Euclidean space, with respect to Minkowski space, by treating $\psib$ as an independent field not related to $\psi$ by Hermitian conjugation. One negative aspect of such construction is that the resulting action is not Hermitian. This might not be a problem for fermions in general, but as we want to preserve the hermiticity property of the matrix-valued bilinear in equation \eqref{eq:matrixB} (true in Lorentzian signature, and needed for our intermediate field analysis), we follow here an alternative route, due mainly to Mehta \cite{Mehta:1986mi} and  van Nieuwenhuizen and Waldron \cite{vanNieuwenhuizen:1996tv}, and which for our purposes boils down to the following observation.
We can define $\psib=\psi^{\rm \dagger} \G_0$ for some matrix $\G_0$ to be determined. 
Requiring $\psib\psi$ and $\psib\g_\m\psi$ to behave as a scalar and a vector, respectively, under rotations allows us to partially fix $\G_0$.
In order to see that, we define the spinorial representation of the generator of rotations as
\be
\Sigma_{\m\n} = -\f14 [\g_\m,\g_\n] =\f{\im}{2}\eps_{\m\n}\g_5 = - \Sigma_{\m\n}^\dagger\;.
\ee
It satisfies
\be
[\Sigma_{\m\n},\g_\r] = (J_{\m\n})_{\r\s} \g_\s\;,
\ee
where
\be
(J_{\m\n})_{\r\s} = \eps_{\m\n} \eps_{\r\s} = \d_{\m\r} \d_{\n\s} -  \d_{\m\s} \d_{\n\r} \;,
\ee
is the vectorial representation of the rotation group, i.e. a rotation is written as
\be
R = e^{\f12 \om_{\m\n} J_{\m\n}}  = \begin{pmatrix} \cos{\a} & \sin{\a} \\ -\sin{\a} & \cos{\a} \end{pmatrix} \;,
\ee
where $\om_{\m\n}=\a \eps_{\m\n}$.
Defining a finite rotation in spinorial representation as 
\be
S(R) =  e^{-\f12 \om_{\m\n} \Sigma_{\m\n} } = e^{\f{\im}{2} \a \g_5 } \neq S^\dagger(R) = S(R^{-1}) \;,
\ee
one can check that
\be
S(R^{-1}) \g_\r S(R) = R_{\r\s} \g_\s \;.
\ee
Note that in the Majorana representation
\be
S(R) = \begin{pmatrix} \cos{\f{\a}{2}} & - \sin{\f{\a}{2}} \\ \sin{\f{\a}{2}} & \cos{\f{\a}{2}} \end{pmatrix} \;,
\ee
is real (thus preserving the reality condition) but non-diagonal, while in the Weyl representation
\be
S(R) = \begin{pmatrix} e^{\im\a/2} & 0 \\ 0 & -e^{\im\a/2} \end{pmatrix} \;,
\ee
is diagonal but complex (as it should be, since there are no real one-dimensional representations of $SO(2)$).
Therefore, for  $\psib\psi$ and $\psib\g_\m\psi$ to transform appropriately we just need that if $\psi\to S(R)\psi$, then $\psib\to \psib S^\dagger(R)$.
This is trivially achieved if $\G_0$ is either the identity or $\g_5$ (or a linear combination of the two), because they both commute with $S^\dagger(R)$.  Either choice can be found in the literature, however, the requirement of a consistent Wick rotation from Lorentzian to Euclidean signature selects $\G_0=\g_5$ (see \cite{Mehta:1986mi,vanNieuwenhuizen:1996tv}\footnote{Our construction for the Majorana case differs both from \cite{Mehta:1986mi,vanNieuwenhuizen:1996tv} and from \cite{Nicolai:1978vc}, which work in four dimensions (where, contrary to two dimensions, there are no real Majorana fermions in Euclidean signature), as well as from \cite{Borthwick:1993pb} (where the rotational  invariance of the reality condition is not preserved).}).

We can also define a continuous chiral transformation as
\be \label{eq:cct1}
\psi \to e^{\theta \g_5} \psi = (\cosh\theta + \g_5 \sinh\theta) \psi \;,
\ee
with real $\theta$. Under such a transformation $\psib\g_\m\psi$ is invariant while $\psib\psi$ and $\psib\g_5\psi$ are not (here we are using $\psib=\psi^{\dagger} \g_5 \to \psib e^{\theta \g_5} $). Furthermore, the combination $(\psib\psi)^2-(\psib\g_5\psi)^2$ is invariant.
However, for no real value of $\theta$ does such a continuous chiral symmetry reduces to the discrete chiral symmetry \eqref{eq:dct}, which is instead obtained for $\theta=\im\pi/2$ (where one should remember that even for complex $\theta$ the transformation for $\psib$ is $\psib \to  \psib e^{\theta \g_5}$, and not $\psib \to  \psib e^{\theta^* \g_5}$).

\section{Bilinears, quadrilinears, and symmetries}
\label{app:invariants}

We denote a bilinear in the fermionic fields by means of parenthesis indicating full contraction of the spinorial indices, e.g.
\be \label{eq:parenthesis}
(\psib_{\bf a} \G \psi_{\bf b}) \equiv \psib_{\bf a}^{\rm \dagger}{}^J  \G_{JK} \psi_{\bf b}^K\; ,
\ee
where $\G$ stands for a generic product of $\g$-matrices.

{\bf Euclidean space symmetry}:\\
In two dimensions we have only 4 independent bilinears, as opposed to 16 in four dimensions:
\begin{align}
& S_{\bf ab} = (\psib_{\bf a} \psi_{\bf b}) \;,\\
& V^\m_{\bf ab} = (\psib_{\bf a} \g^\m \psi_{\bf b}) \;,\\
& P_{\bf ab} = (\psib_{\bf a} \g_5 \psi_{\bf b}) \;,
\end{align}
which behave under the rotation group as scalars, vectors, and pseudoscalars, respectively.
Besides $S_{\bf ab}$, we can form another scalar by inserting a derivative in $V^\m_{\bf ab}$:
\be \label{eq:kin-bil}
(\psib_{\bf a} \slashed{\p} \psi_{\bf b})\;,
\ee
where as usual we defined $\slashed{\p}_\m=\g^\m\p_\m$.

We have only three independent scalars which we can construct out of four fields and with no derivatives:
\begin{align} \label{eq:I_S}
& I_{\bf abcd}^S = S_{\bf ab} S_{\bf cd} \;,\\ \label{eq:I_V}
& I_{\bf abcd}^V = V^\m_{\bf ab} V_\m{}_{\bf cd}  \;,\\ \label{eq:I_P}
& I_{\bf abcd}^P = P_{\bf ab} P_{\bf cd} \;.
\end{align}
Quadrilinears with derivatives and higher order multilinears correspond to non-renormalizable interactions and are therefore not of interest for our purposes.

{\bf Chiral symmetry}:\\
Under the discrete chiral transformation \eqref{eq:dct} all the three quadrilinears \eqref{eq:I_S}-\eqref{eq:I_P} are individually invariant, while among the two scalar bilinears only \eqref{eq:kin-bil} is invariant.

Under the continuous chiral transformations \eqref{eq:cct1}, it's again \eqref{eq:kin-bil} the only invariant bilinear, while among the quadrilinears the invariants are $I_{\bf abcd}^V$ and the combination $I_{\bf abcd}^S-I_{\bf abcd}^P$.

{\bf Flavor symmetry}:\\
In order to complete the construction of the Lagrangian we need to contract the group indices in such a way to build invariants.

\begin{figure}[ht]
 \centering
        \begin{minipage}{0.25\textwidth}
            \centering 
             \includegraphics[scale = 0.75]{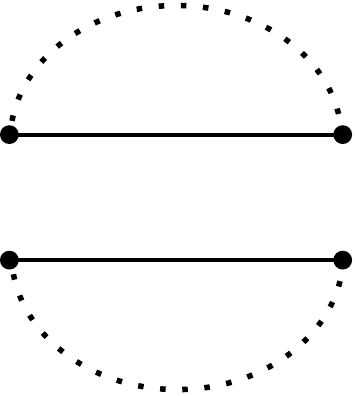}
        \end{minipage}
        \hspace{0.01\textwidth}
        \begin{minipage}{0.25\textwidth}
            \centering
            \includegraphics[scale = 0.65]{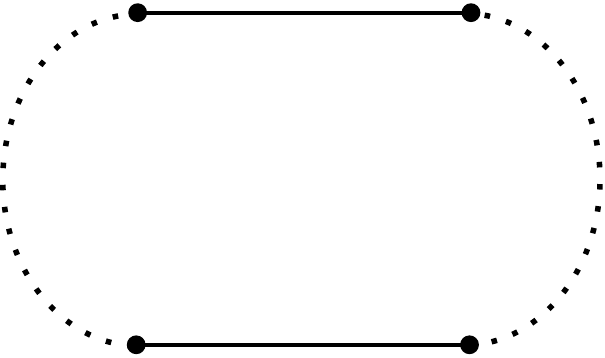}
        \end{minipage}
  \caption{\small{Graphical representation of the interaction vertices of the GN model, displaying the two types of spinor indices contraction (dotted lines): the usual disconnected contraction (left) and the connected one (right).}}
\label{fig:verticesGN}
\end{figure}

In the GN case, i.e. for $G=O(N)$ or $G=U(N)$, there are two possible ways to contract the four indices: one which we will call ``disconnected contraction'', ${\bf a}={\bf b}$ and ${\bf c}={\bf d}$; and one which we will call ``connected contraction'', ${\bf a}={\bf d}$ and ${\bf b}={\bf c}$. The reason for the terminology should be evident from figure \ref{fig:verticesGN}.
However, only one of the two types of contractions is independent.
In fact, we can always use the completeness relation \eqref{eq:compl} to get rid for example of the connected contraction, in favor of a combination of invariants with disconnected contractions \cite{Mitter:1974cy}.
For Majorana fermions $V^\m_{\bf aa}=P_{\bf aa}=0$, hence we have simply $(\psib_{\bf a} \psi_{\bf b}) (\psib_{\bf b} \psi_{\bf a})=-\f12 (\psib_{\bf a} \psi_{\bf a}) (\psib_{\bf b} \psi_{\bf b})$, and thus one unique interaction.
For Dirac fermions we have three independent interactions: $I_{\bf aacc}^S$,  $I_{\bf aacc}^V$, and   $I_{\bf aacc}^P$.

In the tensor case it is well known that there are many ways to construct invariants.\footnote{See for example \cite{Geloun:2013kta} for an enumeration of invariants with $n$ tensors.}
The invariants relevant to our case are discussed in detail in section \ref{sec:models}.

\section{Beta function of the $\l_2$ coupling at leading order}
\label{app:beta2}

In section \ref{sec:SDE-D} we have obtained the exact large-$N$ beta functions for $\l_0^S$ and $\l_1^S$ by means of a Callan-Symanzik equation for the physical mass, and by a simple analysis of the possible diagrams at large $N$.
The same method was not applicable to the $\l_2$ coupling in section \ref{sec:SDE-M}, because it was not possible to solve the SD equations. In this appendix we are going to compute the beta functions for the coupling $\l_2$ of \eqref{eq:general-int-M} and the coupling $\l$ of \eqref{S2_general} with conditions \eqref{eq:I2_cond1}, and show that they are both IR free.

Let us start with the action \eqref{eq:general-int-M}.
Note that, while $\l_0$ only receives quantum corrections from $\l_1^X$, there is no way to protect the latter from radiative corrections generated by $\l_2$. However, the running of the latter is independent of $\l_0$ and $\l_1^X$ at leading order in $1/N$.
In fact, as we explained in figure \ref{fig:beta2}, at leading order in $1/N$ the vertex $I_2$ receives no quantum corrections at all.
Its running can therefore only come from the wave-function renormalization $Z$, but since $I_0$ and $I_1^X$ only contribute to the 2-point function with momentum independent tadpoles, $Z$ must only depend on $\l_2$.
So for the purpose of calculating the running of $\l_2$ we can forget the other couplings.

Since in the presence of $I_2$ the 2-point function needs a multiplicative renormalization, we need to introduce a wave-function renormalization,
\be
\psi = \sqrt{Z} \psi_R \;.
\ee
Rewriting the action in terms of the renormalized fields, the effective coupling is
\be
\l_{2,R} = Z^2 \l_2\;.
\ee
Since at leading order in $1/N$ the part of the 4-point function with $I_2$ structure is exact at  tree level, then $\l_{2,R} $ is the renormalized coupling. Hence, its beta function is
\be \label{eq:beta2-def}
\b_2 = \m\p_\m \l_{2,R} {}_{\big| \l_2\, {\rm fixed}} = \l_{2,R} 2\f{\m\p_\m Z}{Z} \equiv 4\eta  \l_{2,R}\;.
\ee

We can compute $Z$ at lowest order in perturbation theory (and leading order in $1/N$), essentially evaluating $\Sigma$ from our SD equations with the full propagator $G$ replaced by the free propagator. 
It is convenient to write the latter in coordinate space, and use dimensional regularization. 
We have for the propagator:
\be
\begin{split}
C(x) & = \int \f{d^d p}{(2\pi)^d} e^{\im \vec{p}\cdot \vec{x}} \f{-\im \slashed{p}}{p^2}= -\im \g_\m \int \f{d^d p}{(2\pi)^d} e^{\im \vec{p}\cdot \vec{x}} p_\m \int_0^{+\infty} dt\, e^{-t p^2}\\
& = \f{\G(d/2)}{2\pi^{d/2}} \f{\slashed{x}}{x^{d}} \equiv c_d \f{\slashed{x}}{x^{d}}
 \;,
\end{split}
\ee
with $c_2 = 1/(2\pi)$.

The self-energy is
\be
\begin{split}
\Sigma(x) &= -3 \l_2^2 \left(2\,  \g_5 C(x) \g_5 C(-x) \g_5 C(x) \g_5  + \g_5 C(x) \g_5  \Tr[C(-x) \g_5 C(x) \g_5] \right) \\
& = +3 \l_2^2 \m^{4-2d}c_d^3\, \slashed{x} \, \f{2 \slashed{x} \slashed{x}+d_\g x^2}{x^{3d}} \;,
\end{split}
\ee
where $d_\g$ is the dimension of the gamma matrices in $d$ dimensions, and where we introduced the mass scale $\m$ in order to keep $\l_2$ dimensionless.

In order to go back to momentum space we need to compute (here the integral over  $t$ is finite for $d<2$):
\be \label{eq:int_Sigma-x}
\begin{split}
\int d^d x\, & e^{-\im \vec{p}\cdot \vec{x}} \f{x_\m x_\n x_\r}{x^{3d}} 
 = \f{1}{\G(3d/2)} \int d^d x\, e^{-\im \vec{p}\cdot \vec{x}} x_\m x_\n x_\r \int_0^{+\infty} dt \, t^{3d/2-1} e^{-t x^2} \\
& =  -\f{\im \pi^{d/2}}{4\G(3d/2)} \int_0^{+\infty} dt \, t^{d-3} e^{-\f{p^2}{4t}}   \left(\d_{\m\n}p_\r +\d_{\m\r}p_\n +\d_{\n\r}p_\m - \f{p_\m p_\n p_\r}{2 t} \right)\\
& = -\f{\im \pi^{d/2}}{4\G(3d/2)} 4^{2-d} \left( p^{2d-4} \G(2-d) (\d_{\m\n}p_\r +\d_{\m\r}p_\n +\d_{\n\r}p_\m) 
- 2 p^{2d-6} \G(3-d)p_\m p_\n p_\r\right)\\
& = -\im\f{\pi}{8} \Big(\f{1}{\epsilon} -2 \ln (p) +C\Big) (\d_{\m\n}p_\r +\d_{\m\r}p_\n +\d_{\n\r}p_\m)  +\im\f{\pi}{4} \f{p_\m p_\n p_\r}{p^2} +O(\eps)\;,
\end{split}
\ee
where in the last step we used $d=2-\eps$, and $C$ is a finite constant.
Therefore, for the self-energy in momentum space we obtain
\be \label{eq:Sigma-pert}
\begin{split}
\hat\Sigma(p) & = \int d^d x\, e^{-\im \vec{p}\cdot \vec{x}} \Sigma(x) \\
 & =-\im 6 \l_2^2 c_d^3 \f{\pi}{8}  (\g_\m \g_\n \g_\r + \g_\m \d_{\n\r}) \left(\Big(\f{1}{\epsilon}+2 \ln (\m/p) +C\Big)(\d_{\m\n}p_\r +\d_{\m\r}p_\n +\d_{\n\r}p_\m) -\f{p_\m p_\n p_\r}{p^2} \right)\\
 & =-\im 6 \l_2^2 c_2^3 \pi \left(\f{1}{\epsilon}+2 \ln (\m/p) +C'\right) \slashed{p} \;,
\end{split}
\ee
where the finite constant $C$ is replaced by a new constant $C'$ because of the order $\eps$ contributions from $c_d$ and from the $d$-dependence of contraction identities for $\g$ matrices, as well as because we include in it the finite contribution from the last term in the intermediate step.

Plugging \eqref{eq:Sigma-pert} in the SD equation \eqref{eq:SDE-p}, we find 
\be \label{eq:G2-bare}
\hat G(p) = -\im \f{\slashed{p}}{p^2} \left( 1 + 6 \l_2^2 c_2^3 \pi \Big(\f{1}{\epsilon}+2 \ln (\m/p) +C'\Big) +\d Z\right)^{-1} \;,
\ee
where we have included also the counterterm $\d Z=Z-1$.
At this order of perturbation theory we define the wave-function renormalization as (i.e. in a modified minimal subtraction scheme)
\be
Z  = 1 - 6 \l_2^2 c_2^3 \pi \Big(\f{1}{\epsilon} +C'\Big) \;,
\ee
so that the renormalized 2-point function reads
\be \label{eq:G2-ren}
\hat G_R(p) = -\im \f{\slashed{p}}{p^2} \left( 1 + 12 \l_{2,R}^2 c_2^3 \pi \ln (\m/p)  \right)^{-1} \simeq -\im \f{\slashed{p}}{p^2} \left( 1 - 12 \l_{2,R}^2 c_2^3 \pi \ln (\m/p)  \right)\;,
\ee
which has canonical normalization at $p=\m$.

In order to obtain the beta function, we can use the Callan-Symanzik equation for the renormalized 2-point function:
\be
\left( \m\p_\m + \b_2\p_{\l_{2,R}} +2\eta\right)\hat G_R(p) = 0\;.
\ee
From \eqref{eq:beta2-def} we see that the beta function term is higher order in $\l_2$, so it does not contribute to lowest order.
Using \eqref{eq:G2-ren}, we find
\be \label{eq:eta}
\eta= \f{3}{4\pi^2} \l_2^2\;,
\ee
and thus
\be
\b_2 =  \f{3}{\pi^2} \l_2^3 \;.
\ee

The beta function is therefore positive for $\l_2>0$ and negative for $\l_2<0$, hence the coupling is IR free for both signs.

\

In the case of the action  \eqref{S2_general} with conditions \eqref{eq:I2_cond1}, the calculation is essentially as above, but there is only the trace term in the self-energy (multiplied by an extra factor 2 for the two types of vertices). That is, we have
\be
\begin{split}
\Sigma(p) & = \int d^d x\, e^{-\im \vec{p}\cdot \vec{x}} \Sigma(x) \\
 & =-\im 12 \l^2  c_2^3 \f{\pi}{8}  \Big(\f{1}{\epsilon}+2 \ln (\m/p)\Big) \g_\m \d_{\n\r} (\d_{\m\n}p_\r +\d_{\m\r}p_\n +\d_{\n\r}p_\m) +O(\eps^0)\\
 & =-\im 6 \l^2  c_2^3 \pi\Big(\f{1}{\epsilon}+2 \ln (\m/p)\Big) \slashed{p} +O(\eps^0) \;,
\end{split}
\ee
leading again to 
\be
\b_2 =  \f{3}{\pi^2} \l^3 \;.
\ee
%

\section{The other stationary points}
\label{app:extrema}

We solve here the equations of motion for the Dirac model of section \ref{sec:Dirac-1color} in full generality.
To that end, it is convenient to use the representation without the double trace, i.e. with potential
\be \label{eq:L-phi}
V(M,\phi)=\f{1}{2}  \Tr[ M^2] + \frac{\lambda_1 }{2 \pi} \Tr\left[  M^{2} \left( \ln \frac{ 2 \lambda_1  M^{2}}{ \Lambda^2}- 1  \right)\right] 
+ \f{(1 - b ) N}{2} \phi^2 - \sqrt{-b}\, \phi\, \Tr[M]\;.
\ee
The equations of motion read
\be \label{eq:mu-eom-phi1}
\f{\p V}{\p M_{ij}}=0  \to \d_{ij} \left( \m_i  + \f{\l_1}{\pi} \m_i \ln\f{2\l_1\m_i^2}{\L^2} - \sqrt{-b}\, \phi \right)=0\;,
\ee
\be \label{eq:mu-eom-phi2}
\f{\p V}{\p \phi}=0  \to \Tr[M]=\f{1-b}{\sqrt{-b}} N \phi\;.
\ee
The latter is simply a constraint relating the new variable $\phi$ with the trace of $M$.
The first equation can be simplified by a rescaling of the eigenvalues, defining
\be
\m_i^2 = \f{\L^2}{2\l_1} e^{-\f{\pi}{\l_1}}x_i^2\;.
\ee
We obtain
\be
x_i \ln |x_i| = \f{\pi\sqrt{2}}{\L\sqrt{\l_1}} e^{\f{\pi}{2\l_1}} \sqrt{-b}\phi \equiv z\;,
\ee
which is solved (almost by definition) by
\be
x_i = \pm e^{W(\pm z)}\;,
\ee
where $W(z)$ is the Lambert-W function (or product logarithm).
Denoting by $n_+=N_+/N$ the fraction of eigenvalues with the plus sign, and plugging the solution of \eqref{eq:mu-eom-phi1} into \eqref{eq:mu-eom-phi2}, we arrive at a self-consistency equation determining $z$ as a function of $n_+$:
\be \label{eq:consistency-z}
n_+ e^{W(z)} - (1-n_+) e^{W(-z)} = - \f{2(1-b)\l_1}{b\pi} z \;.
\ee
Notice that the real branch of the Lambert-W function is defined for $z\geq -1/e$, and since for $0<n_+<1$ we have both signs in its argument for the solutions above, we have the constraint $|z|\leq 1/e$.

For $n_+ =1$ we find (using $e^{W(z)}=z/W(z)$)
\be
W(z) = -\f{b\pi}{2\l_1(1-b)}\;,
\ee
in agreement with the symmetric solution \eqref{eq:sym-mu}.
However, in the range $|z|\leq 1/e$, such equation can only be satisfied for
\be
-\f{2\l_1W(1/e)}{\pi-2\l_1W(1/e)} \leq b \leq \f{2\l_1}{\pi+2\l_1} =b_c\;,
\ee
hence the symmetric solution becomes disconnected from this family of solutions outside of this range of $b$ (notice that for $\l_1>\pi/(2W(1/e))$ the lower bound is replaced by $b>-\infty$).

Solving equation \eqref{eq:consistency-z} for $z$ is difficult, it is easier instead to solve it for $n_+$.
Since $W(0)=0$, for the special case $z=0$ (for which $\m_i$ reduces to \eqref{eq:broken-mu}), we find $n_+=1/2$. That is, we find the traceless solution discussed in section \ref{sec:stability-Dirac} as a special case of the more general class of solutions.
For $z\neq 0$, we obtain
\be
n_+ = \f{(\pi b + 2\l_1 (1-b) W(-z))W(z)}{\pi b (W(z)-W(-z))}\;.
\ee

The computation of the Hessian for the general case is frustrated by the non-commutativity of both $M$ and $M^2$ with the infinitesimal variation $dM$.
We can however check that the value of the potential at these other solutions (with the constraint $0\leq n_+\leq 1$),
\be
V(M(n_+)) = \L^2 e^{-\f{\pi}{\l_1}} z^2\f{(\pi b + 2\l_1 (1-b) (W(z)+W(-z))}{4\pi^2 b W(z)W(-z))}\;,
\ee
always lies between $v_1$ and $v_2$, equation \eqref{eq:v1} and \eqref{eq:v2}, respectively. Therefore, whether they are local minima, maxima or saddles, they do not affect the identification of the global minimum.
Lastly, at $b=0$ they all have the same value of the potential, i.e. they are the $N+1$ degenerate solutions discussed below equation \eqref{eq:broken-mu}.

\providecommand{\href}[2]{#2}\begingroup\raggedright\endgroup


\end{document}